\documentclass[pra,onecolumn,floatfix,a4paper,superscriptaddress]{revtex4}
\usepackage{bm,color,graphicx,amsmath,txfonts}

\usepackage[colorlinks, citecolor=blue,linkcolor=blue]{hyperref}

\begin{document}

	\title{Kerr-induced nonreciprocal transparency and group delay in a hybrid cavity magnomechanical system}

	\author{M. Amghar}
	\affiliation{LPTHE-Department of Physics, Faculty of sciences, Ibnou Zohr University, Agadir, Morocco}
	
	\author{N. Chabar}
	\affiliation{LPTHE-Department of Physics, Faculty of sciences, Ibnou Zohr University, Agadir, Morocco}
	
	\author{M. Amazioug}\email{m.amazioug@uiz.ac.ma}
	\affiliation{LPTHE-Department of Physics, Faculty of sciences, Ibnou Zohr University, Agadir, Morocco}
	
	\author{R. Altuijri}
	\affiliation{Department of Physics, College of Science, Princess Nourah bint Abdulrahman University, P. O. Box 84428, Riyadh 11671, Saudi Arabia}
	
	\author{A. N. Al-Ahmadi}
	\affiliation{Department of Physics, College of Sciences, Umm Al-Qura University, Makkah 24382, Saudi Arabia}
	
	\author{R. Ahl Laamara}
	\affiliation{LPHE-Modeling and Simulation, Faculty of Sciences, Mohammed V University in Rabat, Rabat, Morocco}
	
	\author{A. Abdel-Aty}
	\affiliation{Department of Physics, College of Sciences, University of Bisha, Bisha, 61922, Saudi Arabia}
	
	\begin{abstract}
		
		We propose a scheme for realizing nonreciprocal transparency, Fano resonances, and slow/fast light in a hybrid cavity magnomechanical system containing two YIG spheres and a mechanical resonator. The nonreciprocal behavior originates from the magnon Kerr nonlinearity, which induces direction-dependent frequency shifts and modifies the interference pathways among cavity photons, magnons, and phonons. We show that the hybrid system supports multiple transparency windows arising from magnon- and magnomechanical-induced interference processes. The Kerr interaction strongly reshapes these transparency features, producing asymmetric Fano resonance line shapes and enabling controllable nonreciprocal transmission. Furthermore, the associated dispersion exhibits pronounced directional asymmetry, leading to giant differences in the group delay for opposite propagation directions and allowing reversible switching between slow- and fast-light regimes. We investigate the roles of hybrid coupling strengths and dissipation channels and identify parameter regimes where the nonreciprocal response is maximized. These findings establish Kerr-engineered magnomechanical systems as promising platforms for integrated nonreciprocal microwave photonics and quantum information technologies.
		
	\end{abstract}
	
	\date{\today}
	
	\maketitle
	\section{INTRODUCTION}
\begin{figure}[t]
		\centering
		\includegraphics[scale=0.33]{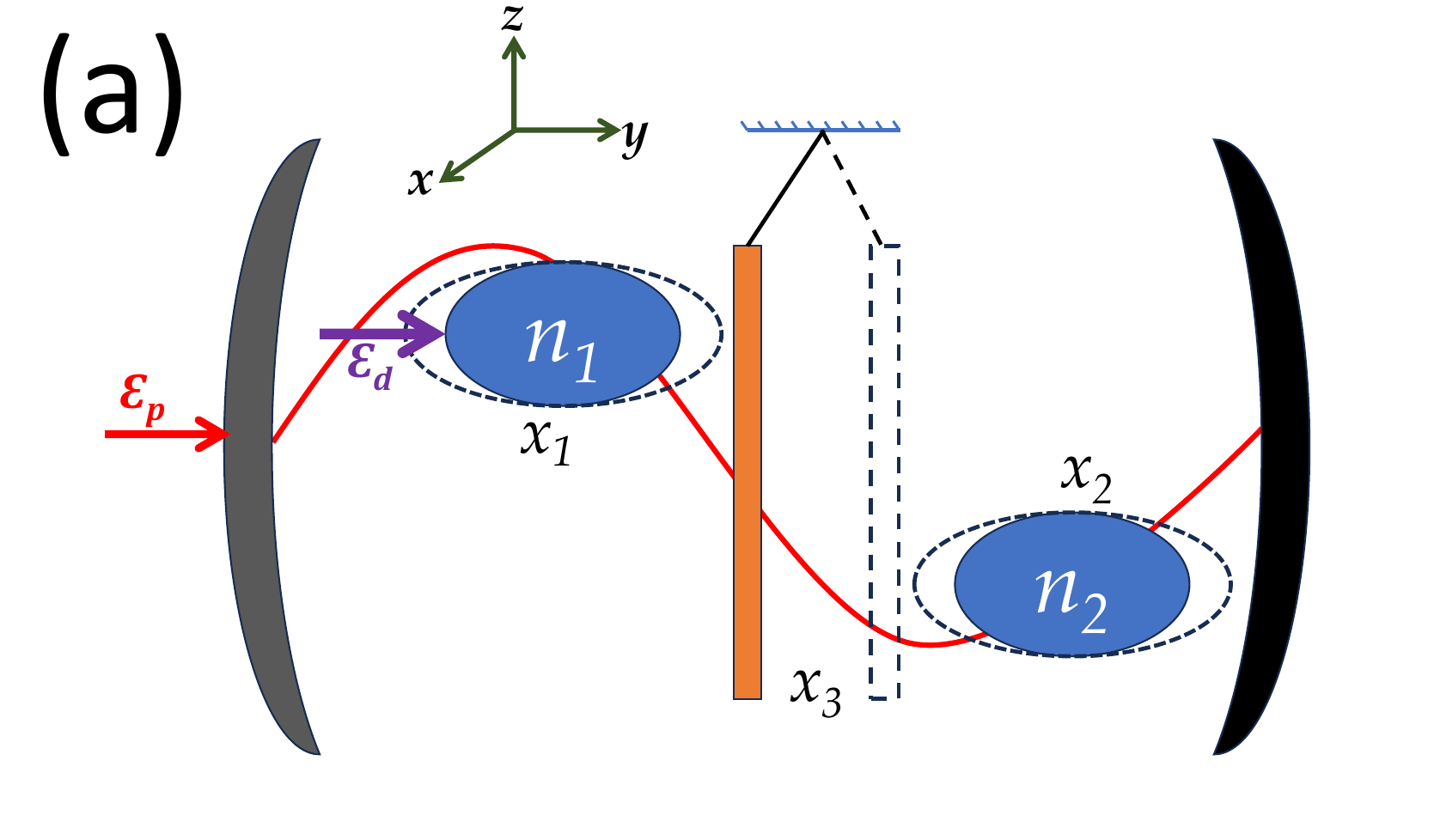}
		\includegraphics[scale=0.3]{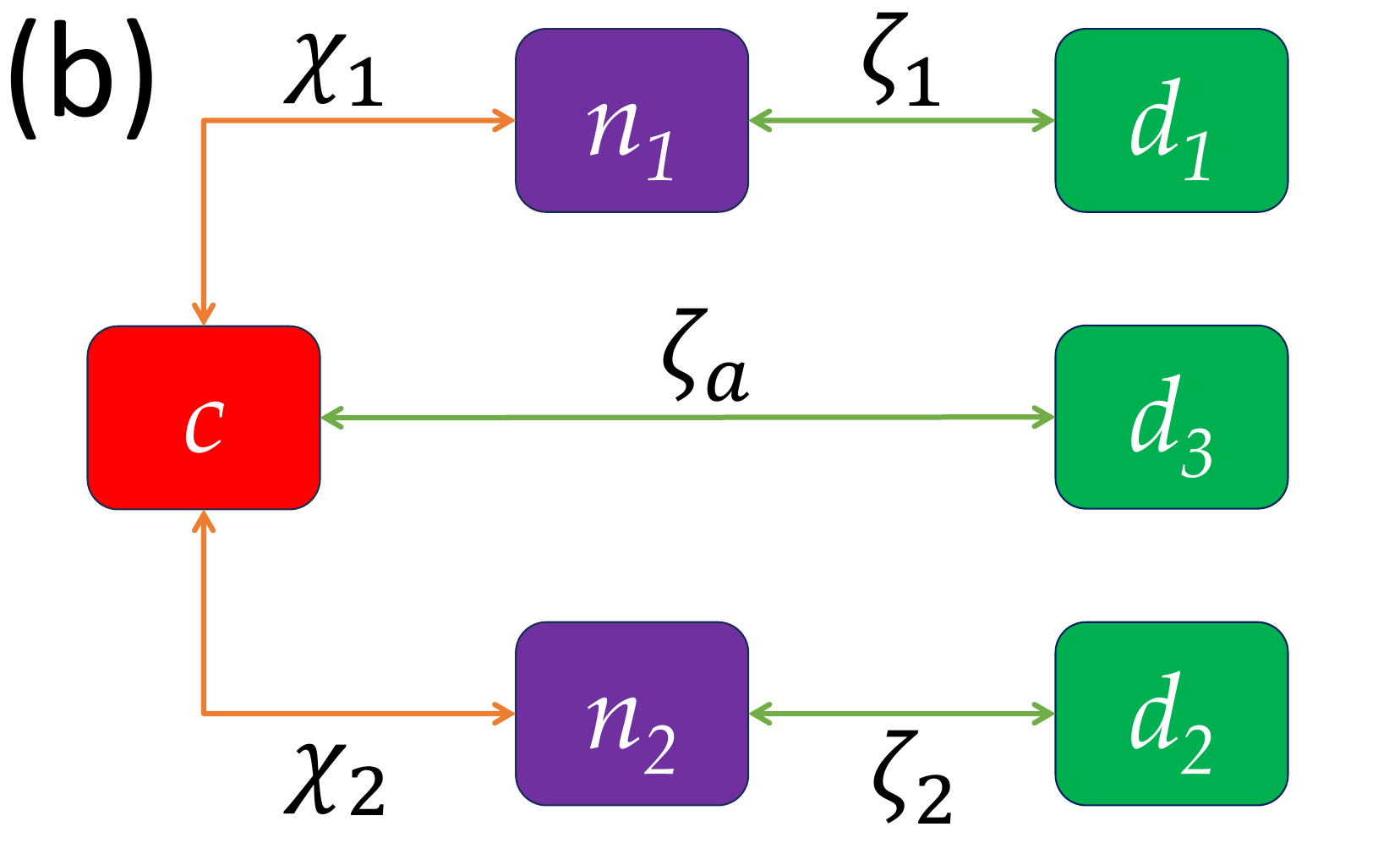}
		\caption{(a) Schematic illustration of the proposed hybrid cavity magnomechanical architecture. The system consists of two yttrium iron garnet (YIG) spheres together with a mechanical membrane placed inside a microwave cavity, which is driven by a weak probe field at frequency $\omega_p$. A static magnetic field applied along the $z$-axis generates collective magnon excitations in the YIG spheres, which are strongly coupled to the cavity electromagnetic mode. Simultaneously, the magnetic bias field activates magnetostrictive effects, leading to effective magnon-phonon coupling and enabling coherent hybrid interactions among photonic, magnonic, and mechanical degrees of freedom. In addition, the first sphere ($n_1$), which exhibits Kerr nonlinearity, is driven by an external microwave control field applied along the $y$-direction. (b) Schematic representation of the coupling map showing interactions among cavity photons, magnons, and mechanical vibrations in the system.}\label{fig1}
	\end{figure}

	Cavity magnomechanics \cite{0} has recently emerged as a powerful platform for investigating coherent interactions among electromagnetic, magnetic, and mechanical excitations in hybrid quantum systems. Owing to their intrinsic nonlinearities, long coherence times, and high tunability, these systems offer promising opportunities for both fundamental research and applications such as quantum information processing, coherent signal transduction, microwave photonics, and precision sensing. In particular, microwave cavity architectures provide a versatile and experimentally accessible framework for engineering photon–magnon–phonon coupling. Among the various magnetic materials, yttrium iron garnet (YIG) stands out due to its ultralow magnetic damping, strong spin–photon coupling strength, and long magnon lifetime \cite{1,2,3}. In a typical configuration, a YIG sphere embedded in a microwave cavity enables the interaction of collective spin excitations (magnons) with cavity photons and mechanical vibrations, giving rise to a wide range of nonlinear and quantum interference effects. Recent studies have demonstrated phenomena such as squeezed-state generation \cite{4}, phase-controlled microwave transmission \cite{5}, optical bistability \cite{6}, dynamical backaction \cite{7}, exceptional-point physics \cite{8}, coherent state transfer \cite{9}, strong magnon–microwave coupling \cite{10}, nonreciprocal quantum phase transitions \cite{11}, cavity-mediated microwave effects \cite{12,13,14,15}, hybrid entanglement generation \cite{12,16} and magnomechanically induced transparency \cite{17,18,19}, highlighting the potential of these systems for tunable quantum functionalities and integrated microwave technologies. Additional related works can be found in Refs.~\cite{21,22,23}. Among these phenomena, MMIT has attracted particular attention because it originates from destructive interference between different excitation pathways mediated by the magnon-phonon interaction \cite{18,24,25}. Analogous to electromagnetically induced transparency in atomic systems \cite{26,27,28}, MMIT is characterized by the appearance of a narrow transparency window within an absorption spectrum, accompanied by steep dispersion and enhanced group delay \cite{29}. These features enable efficient control of microwave signal propagation and have stimulated extensive interest in applications such as coherent information storage \cite{30,31}, tunable microwave filtering \cite{32}, slow- and fast-light manipulation \cite{33}, and quantum state engineering, establishing cavity magnomechanical systems as versatile platforms for reconfigurable microwave photonic and quantum technologies \cite{34,35,36}.\\ \\
	Nonreciprocal wave propagation, characterized by direction-dependent transmission properties, has become a central topic in modern photonics and hybrid quantum systems. By allowing electromagnetic signals to propagate preferentially in one direction while suppressing backward transmission, nonreciprocal devices play a crucial role in eliminating undesired back reflections and enabling robust signal routing. Such capabilities are essential for quantum communication, information processing, and the realization of large-scale quantum networks \cite{37,38}. Among the various approaches proposed to achieve nonreciprocity, the Sagnac effect induced by spinning resonators has emerged as a particularly efficient mechanism for breaking the symmetry between forward and backward propagation. This approach has enabled the realization of a wide range of nonreciprocal phenomena, including directional entanglement and quantum steering \cite{39,40,41}, nonreciprocal squeezing \cite{42}, and asymmetric photon blockade \cite{43,AdP25,Choas26}. Moreover, recent experiments have demonstrated Sagnac-induced nonreciprocal responses in both optical and microwave platforms \cite{44}, highlighting the potential of rotationally induced effects for engineering nonreciprocal functionalities in hybrid quantum devices. \\ \\ 
	In addition to rotation-based approaches, nonlinear magnonic interactions provide another powerful route for achieving nonreciprocal behavior. In particular, the rotation of the YIG sphere generates an effective magnetic field \cite{PRL, Nature, EPJP, ArXiv}. Also, the magnon Kerr nonlinearity can generate asymmetric responses when the direction of the external bias magnetic field is reversed \cite{45}. Such Kerr-induced nonreciprocity has already enabled the experimental realization of directional tripartite entanglement in cavity magnonic systems \cite{46}. More broadly, nonlinear Kerr effects offer highly tunable mechanisms for manipulating microwave transmission, quantum correlations \cite{47}, and MMIT in hybrid platforms \cite{48}. These developments open new opportunities for designing re-configurable nonreciprocal devices and exploring controllable quantum phenomena in macroscopic cavity magnomechanical systems.\\ \\
	Asymmetry induced by the Kerr nonlinearity in MMIT plays a crucial role in shaping the spectral response of hybrid cavity systems, which gives rise to Fano resonance phenomena. The Fano resonance is a fundamental interference effect closely related to induced transparency, first reported in atomic systems \cite{49}. It originates from the quantum interference between a discrete resonant channel and a continuum of excitation pathways, leading to highly asymmetric spectral line shapes with pronounced minima in the absorption spectrum \cite{50}. Owing to its strong sensitivity to interference conditions, Fano resonance has been widely studied in diverse physical platforms, including photonic crystals \cite{51}, coupled micro-resonator systems \cite{52}, and optomechanical configurations \cite{53}. More recently, Fano-like asymmetric spectral features have been observed in hybrid cavity magnomechanical systems \cite{0}, arising from the interplay of magnon–photon–phonon interactions. This Kerr-induced asymmetry is closely connected to the asymmetric response of MMIT and nonreciprocal transmission, highlighting the role of interference engineering in controlling wave propagation in hybrid cavity systems.\\ \\
	In this work, we investigate the nonreciprocal behavior of magnomechanically induced transparency (MMIT), Fano resonances, and slow/fast light propagation in a cavity magnomechanical system under the influence of the Kerr effect. The system consists of two high-quality yttrium iron garnet (YIG) spheres and a mechanical membrane embedded in a microwave cavity (see Figure \ref{fig1}). This setup builds upon our previous investigation of a similar membrane-in-the-middle configuration, where nonreciprocal transparency windows, Fano resonances, and slow/fast light were instead induced via the Barnett effect \cite{ArXiv}. Here, we pivot our focus to analyze the role of the Kerr nonlinearity together with photon–phonon–magnon interactions on the absorption and dispersion spectra. In addition, we discuss the emergence of Fano resonances in the output field. We then examine the influence of the cavity decay rate and magnon dissipation rate on the MMIT response. Furthermore, we investigate slow and fast light propagation and show that the group delay can be controlled via the tunability of the interaction strengths and the sign of the Kerr nonlinearity.\\

	The remainder of this paper is organized as follows. In Subsec.~\ref{010}, we describe the proposed hybrid system and derive its Hamiltonian. In Subsec.~\ref{011}, we derive the corresponding quantum Langevin equations (QLE) and obtain the analytical expression for the output field. In Subsec. \ref{xx}, we analyze magnomechanically induced transparency and investigate the effects of photon–phonon–magnon coupling strengths, Kerr nonlinearity, as well as cavity decay and magnon dissipation rates on the output spectrum. In Subsec. \ref{xxx}, we study the group delay associated with slow and fast light propagation. In Sec. \ref{xxxx}, we examine the nonreciprocal behavior of both absorption and group delay under different system parameters. In Sec. \ref{Fea}, we discuss the experimental feasibility of the proposed scheme using experimentally accessible parameters and realistic system configurations. Finally, Sec. \ref{xxxxx} is devoted to the concluding remarks.
	\section{Theoretical Model}\label{01}
	\subsection{System Hamiltonian}\label{010}
   Figure~\ref{fig1} schematically illustrates a hybrid cavity magnomechanical system consisting of two ferromagnetic YIG spheres and a mechanical membrane positioned inside of the microwave cavity. The magnon modes of the two YIG spheres, denoted by $n_1$ and $n_2$, are excited by an external static magnetic field applied along the $z$-axis and are coupled to the cavity field via the magnetic-dipole interaction. The mechanical vibrations of the YIG spheres, generated by magnetostrictive forces, are described by the phonon modes $d_1$ and $d_2$, which give rise to the magnon-phonon interaction. The mechanical membrane supports the phonon mode $d_3$, which couples to the cavity field through the photon-phonon interaction. The system is driven by a strong control field applied to the first YIG sphere ($n_1$), while a weak probe field is injected into the cavity. Under the rotating-wave approximation and in a reference frame rotating at the driving-field frequency $\omega_d$, the Hamiltonian of the hybrid cavity magnomechanical system takes the form (with $\hbar=1$)
	\begin{equation}\label{H}
		\begin{aligned}
			H = & \, \Delta_c c^{\dagger} c+\Delta_{n_1} n_1^{\dagger} n_1+\Delta_{n_2} n_2^{\dagger}n_2 +\sum_{r=1}^{3} \frac{\omega_{d_r}}{2} \left(y_r^2 + x_r^2 \right)\\
			&+\sum_{r=1,2} \left[ \chi_r \left(c n_r^\dagger + c^\dagger n_r \right) + \zeta_{0r} n_r^\dagger n_r x_r \right]-\chi_cc^\dagger c x_3\\
			& +K\left(n_1^\dagger n_1\right)^2+i \Omega\left(n_1^{\dagger} n_1-\text { H.c }\right)\\
			&+i\left(c^{\dagger} \epsilon_p e^{-i \delta t}-\text { H.c }\right),
		\end{aligned}
	\end{equation}
	where $\Delta_c=\omega_c-\omega_d$, $\Delta_{n_1}=\omega_{n_1}-\omega_d$, $\Delta_{n_2}=\omega_{n_2}-\omega_d$, and $\delta=\omega_p-\omega_d$. Here, the parameters $\omega_c$, $\omega_{n_1}$, and $\omega_{n_2}$ denote the resonance frequencies of the cavity and magnon modes, respectively. The first term in Eq.~\eqref{H} represents the free Hamiltonian of the cavity mode, where $c^\dagger$ and $c$ denote the creation and annihilation operators, satisfying the bosonic commutation relation $[c,c^\dagger]=1$. The second and third terms represent the free Hamiltonians of the magnon modes, where $n_r^\dagger$ and $n_r$ $(r=1,2)$ denote the corresponding creation and annihilation operators. These operators describe the collective spin excitations obtained through the Holstein-Primakoff transformation and satisfy the bosonic commutation relations $[n_r,n_r^\dagger]=1$. The magnon frequency can be conveniently controlled by adjusting the bias magnetic field $H_d$ according to the relation $\omega_n=\gamma H_d$, where $\gamma/2\pi=28,\text{GHz/T}$ denotes the gyromagnetic ratio. The fourth term represents the free Hamiltonian of the phonon modes, where $x_r$ and $y_r$ $(r=1,2,3)$ denote the dimensionless position and momentum quadratures of each mechanical mode with frequencies $\omega_{d_r}$, satisfying the canonical commutation relation $[x_r,y_r]=i$. The fifth term describes the interaction between the cavity mode and the magnon modes, characterized by the cavity-magnon coupling strength $\chi_r$. The sixth term in Eq.~\eqref{H} describes the interaction between the magnon and mechanical modes, with $\zeta_{0r}$ denoting the single-magnon-phonon coupling strength arising from the magnetostrictive effect. The seventh term describes the photon-phonon interaction between the cavity mode and the membrane mode $d_3$, characterized by the single-photon coupling strength $\chi_c$. The eighth term represents the magnon self-Kerr nonlinearity, characterized by the self-Kerr coefficient $K$, which gives rise to an intensity-dependent frequency shift of the magnon mode. The coefficient is given by $K=\mu k_\text{am}\gamma^2/(M^2V_\text{YIG})$, where $\mu$ is the magnetic permeability of free space, $k_\text{am}$ is the first-order anisotropy constant of the YIG sphere, and $M$ is the saturation magnetization. The strength of the self-Kerr nonlinearity increases as the diameter of the YIG sphere decreases. Moreover, $K$ can be positive or negative depending on the alignment of the magnetic field along the crystallographic axes [100] and [110], respectively. Its magnitude can be tuned from $0.05\text{nHz}$ to $100\text{nHz}$ by varying the YIG sphere diameter from $1\text{mm}$ to $100\mu\text{m}$. The ninth term describes the interaction between the microwave driving field and the first magnon mode, where the Rabi frequency $\Omega=\frac{\sqrt{5N}}{4}\,\gamma_GH_d$ characterizes the coupling strength. Here, $N=\rho V_\text{YIG}$ is the total number of spins in the YIG sphere, with $\rho$ and $V_\text{YIG}$ denoting the spin density and the volume of the YIG sphere, respectively, and $H_d$ is the amplitude of the driving magnetic field. Finally, the last term describes the interaction between the cavity mode and the weak probe field, where the probe-field amplitude is given by $\epsilon_p=\sqrt{2k_c\mathcal{P}/\hbar\omega_p}$, with $k_c$ denoting the cavity decay rate and $\mathcal{P}$ ($\omega_p$) representing the power (frequency) of the input microwave probe field.
	\subsection{The quantum dynamics and the output field}\label{011}
    In this subsection, we investigate the quantum dynamics of the hybrid cavity magnomechanical system using the Heisenberg-Langevin formalism. By including the effects of dissipation and quantum fluctuations, the quantum Langevin equations for the system can be written as follows
	\begin{equation}\label{a}
		\begin{aligned}
			&\begin{aligned}
				& \dot{c}=-(\kappa_c+i\Delta_c)c-i \sum_{r=1,2}\chi_rn_r+i\chi_ccx_3+\epsilon_p e^{-i\delta t}+\sqrt{2\kappa_c} c^{in}, \\
				& \dot{n_1}=-(\kappa_{n_1} +i \Delta_{n_1} )n_1-i \chi_1c-i\zeta_{01}n_1x_1-2iKn_1^\dagger n_1n_1+\Omega+\sqrt{2 \kappa_{n_1}} n_1^{i n}, \\
				& \dot{n_2}=-(\kappa_{n_2} +i \Delta_{n_2} )n_2-i \chi_2c-i\zeta_{02}n_2x_2+\sqrt{2 \kappa_{n_2}} n_2^{i n}, \\
				&\dot{x_r}  =\omega_{d_r} y_r,\quad \dot{y_r}=-\omega_{d_r} x_r-\zeta_{0r}n_r^\dagger n_r-\gamma_{d_r}y_r+\xi_r, \quad \text{with}\quad r=1,2\\
				&\dot{x_3}  =\omega_{d_3} y_3,\quad \dot{y_3}=- \omega_{d_3} x_3+\chi_c c^\dagger c-\gamma_{d_3}y_3+\xi_3, \\
			\end{aligned}\\
		\end{aligned}
	\end{equation}
	where $\kappa_{c}$, $\kappa_{n_r}$, and $\gamma_{d_r}$ ($\gamma_{d_3}$) denote the decay rates of the cavity, magnon, and phonon modes, respectively. Here, $c^{\mathrm{in}}$ and $n_r^{\mathrm{in}}$ denote the corresponding input noise operators, while $\xi_r$ corresponds to the Brownian noise acting on the mechanical modes. These noise operators have zero mean values, i.e., $\langle v \rangle = 0$ with $v = c^{\mathrm{in}}, n_1^{\mathrm{in}}, n_2^{\mathrm{in}}, \xi_1, \xi_2, \xi_3$ \cite{24}. To investigate the response of the system to the weak probe field, we employ the standard linearization procedure. Each system operator in Eq.~\eqref{a} is expressed as the sum of its steady-state mean value and its fluctuation, i.e., $\langle \mathcal{Y} \rangle = \mathcal{Y}_s + \mathcal{Y}_- e^{-i \delta t} + \mathcal{Y}_+ e^{i \delta t}$, where $\mathcal{Y} = {c, n_1, n_2, x_1, x_2, x_3, y_1, y_2, y_3}$ \cite{54,55}. Substituting these expressions into Eq.~\eqref{a} and retaining only the steady-state terms, we obtain
	\begin{equation}
		\begin{aligned}
			&  c_{s}=\frac{-i\chi_1n_{1s}-i\chi_2n_{2s}}{\kappa_c+i\bar{\Delta}_c},\quad n_{1s}=\frac{\Omega-i \chi_1 c_{s}}{\kappa_{n_1}+i\left( \bar{\Delta}_{n_1}+\Delta K\right)}, \\
			&n_{2s}=\frac{-i \chi_{2} c_{s}}{\kappa_{n_2}+i \bar{\Delta}_{n_2}},\quad x_{1s}=\frac{-\zeta_{0 1}|n_{1s}|^2}{\omega_{d_1}}, \\
			& x_{2s}=\frac{-\zeta_{0 2}|n_{2s}|^2}{\omega_{d_2}},\quad x_{3s}=\frac{\chi_c|c_{s}|^2}{\omega_{d_3}},
		\end{aligned}
	\end{equation}
	where $\bar{\Delta}_c = \Delta_c - \chi_c x_{3s}$, $\bar{\Delta}_{n_1} = \Delta_{n_1} + \zeta_{01} x_{1s}$, $\bar{\Delta}_{n_2} = \Delta_{n_2} + \zeta_{02} x_{2s}$, and $\Delta K=2K|n_{1s}|^2$ denote the effective detunings of the cavity, the magnon modes, and the Kerr-induced frequency shift, respectively. Since the microwave driving field is much stronger than the weak probe field $\epsilon_p$, the linearized quantum Langevin equations can be solved perturbatively to first order in $\epsilon_p$, while higher-order terms are neglected. As a result, the solution for the cavity mode is given by (see Appendix~\ref{AP})
	\begin{equation}\label{o}
		c_-=\epsilon_p\times\left(\alpha_1+\frac{i\chi_1\mathcal{J}}{\mathcal{I}}+\frac{i\chi_2\mathcal{S}}{\mathcal{R}}-\frac{\mathcal{G}_{c}\mathcal{W}}{\mathcal{V}}\right)^{-1}.
	\end{equation}
	Here, $\mathcal{G}_c = \chi / \sqrt{2}$, with $\chi = i \sqrt{2} \chi_c c_s$ representing the effective optomechanical coupling strengths. It should be noted that the solution for the cavity in Eq.~\eqref{o} is obtained under the condition $|\bar{\Delta}_c|, |\bar{\Delta}_{n_1}|, |\bar{\Delta}_{n_2}| \gg \kappa_c, \kappa_{n_1}, \kappa_{n_2}$.	
	Finally, using the standard input–output relation \(\epsilon_{\mathrm{out}} = \epsilon_{\mathrm{in}} - 2 \kappa_c \langle c \rangle\) \cite{Walls1994}, the amplitude of the output field is
	\begin{equation}\label{E}
		\epsilon_{\mathrm{out}} = \frac{2\kappa_c c_-}{\epsilon_p} = \text{Re}[\epsilon_{\mathrm{out}}] + i \text{Im}[\epsilon_{\mathrm{out}}],
	\end{equation}
	where \(\text{Re}[\epsilon_{\mathrm{out}}]\) and \(\text{Im}[\epsilon_{\mathrm{out}}]\) correspond to the absorption and dispersion spectra of the probe field, respectively. The rescaled transmission field corresponding to the probe field, as given by Eq.~\eqref{E}, can be expressed as follows
	\begin{equation}
		T=1+\epsilon_{out}.
	\end{equation}
	The phase $\phi$ of the transmitted probe field $T$ is defined as follows
	\begin{equation}\label{T}
		\phi=\text{Arg}[T].
	\end{equation}
	\section{RESULTS AND DISCUSSION}
	\subsection{Magnomechanically induced transparency}\label{xx}
	In this subsection, we investigate magnomechanically induced transparency in the proposed hybrid cavity magnomechanical system and analyze the influence of the cavity, magnon, and mechanical decay rates on the transparency windows. The parameters employed in our study are based on recent experimental implementation of a hybrid magnomechanical system \cite{0,24}: $\omega_c/2 \pi=10$ GHz, $\omega_{d}/2 \pi=\omega_{d_1}/2 \pi=\omega_{d_2}/2 \pi=\omega_{d_3}/2 \pi=10$ MHz, $\gamma_{d_1}/2 \pi=\gamma_{d_2}/2 \pi=\gamma_{d_3}/2 \pi=100$ Hz, $\omega_{n_1,n_2}/2 \pi=10$ GHz, $\kappa_c/2 \pi=2.1$ MHz, $\kappa_{n_1}/2 \pi=\kappa_{n_2}/2 \pi=0.1$ MHz, $\chi_1/2\pi=\chi_2/2\pi=1.5$ MHz, $\Delta_c=\omega_{d_1}$, $\Delta_{n_r}=\omega_{d_1}$ ($r=1,2$ ) and $\omega_{d}/2\pi=10$ GHz. The amplitude of the driving magnetic field is taken as $B_{1}=3.6\times10^{-5}$ T, corresponding to a driving power of $P_{1}=7.6$ mW and a single-magnon-phonon coupling strength of $R_{01}/2\pi=0.2$ Hz for a YIG sphere with a diameter of $250~\mu$m. For these parameters, the total number of spins is $N\simeq3.5\times10^{16}$, corresponding to a steady-state magnon population of $|\langle n_{1(2)}\rangle|\simeq1.1\times10^7$. Consequently, $\langle n_{1(2)}^\dagger n_{1(2)}\rangle\simeq1.2\times10^{14}\ll5N=1.8\times10^{17}$, confirming the validity of the Holstein-Primakoff approximation. \\
	\begin{figure}
		\begin{center}
			\includegraphics[scale=0.43]{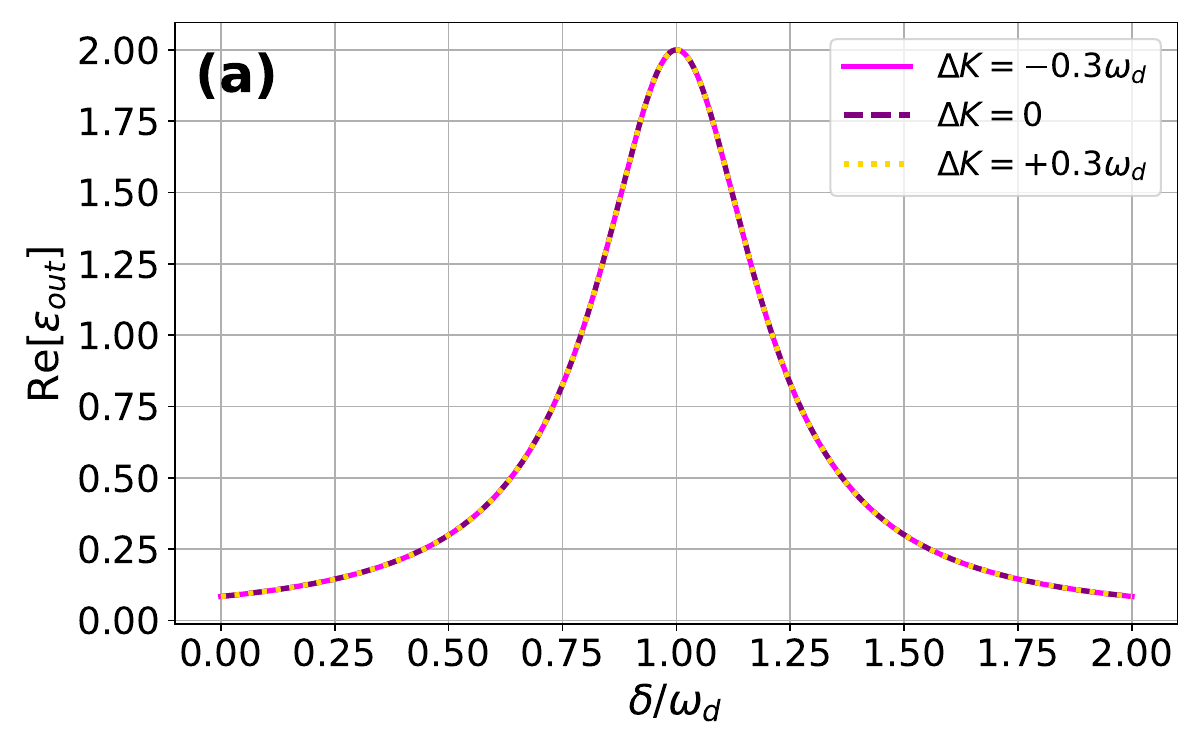}
			\includegraphics[scale=0.43]{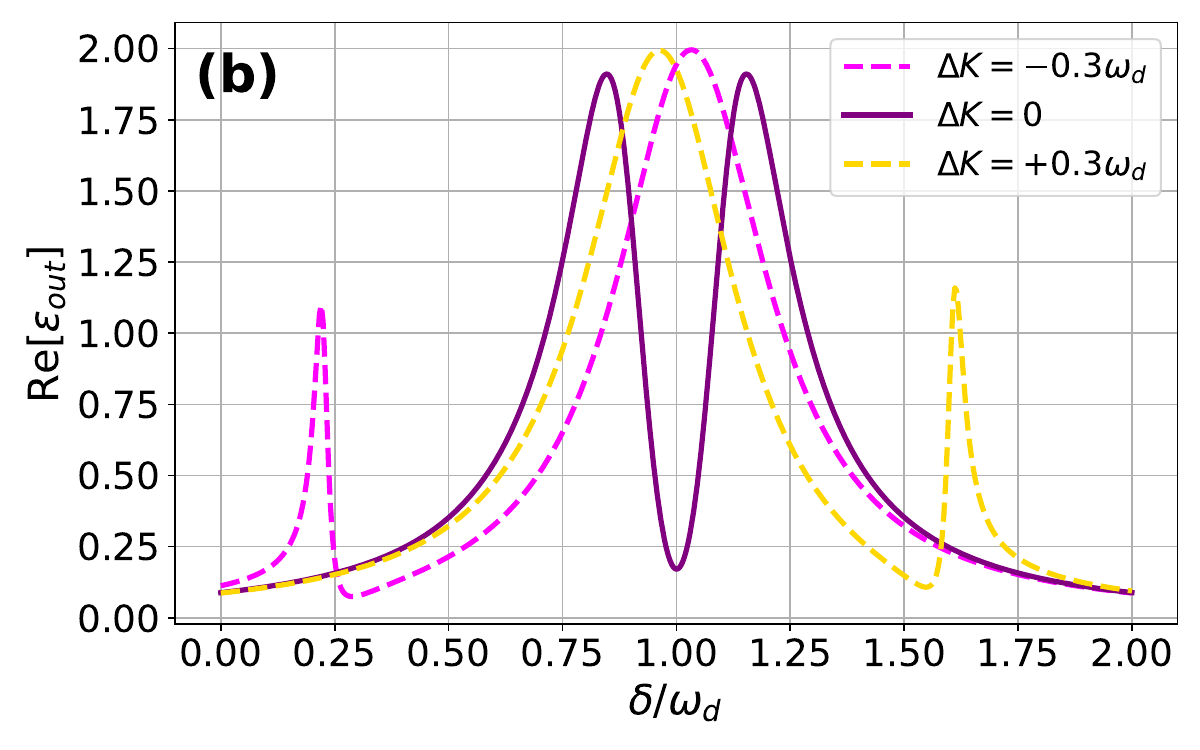}
			\includegraphics[scale=0.43]{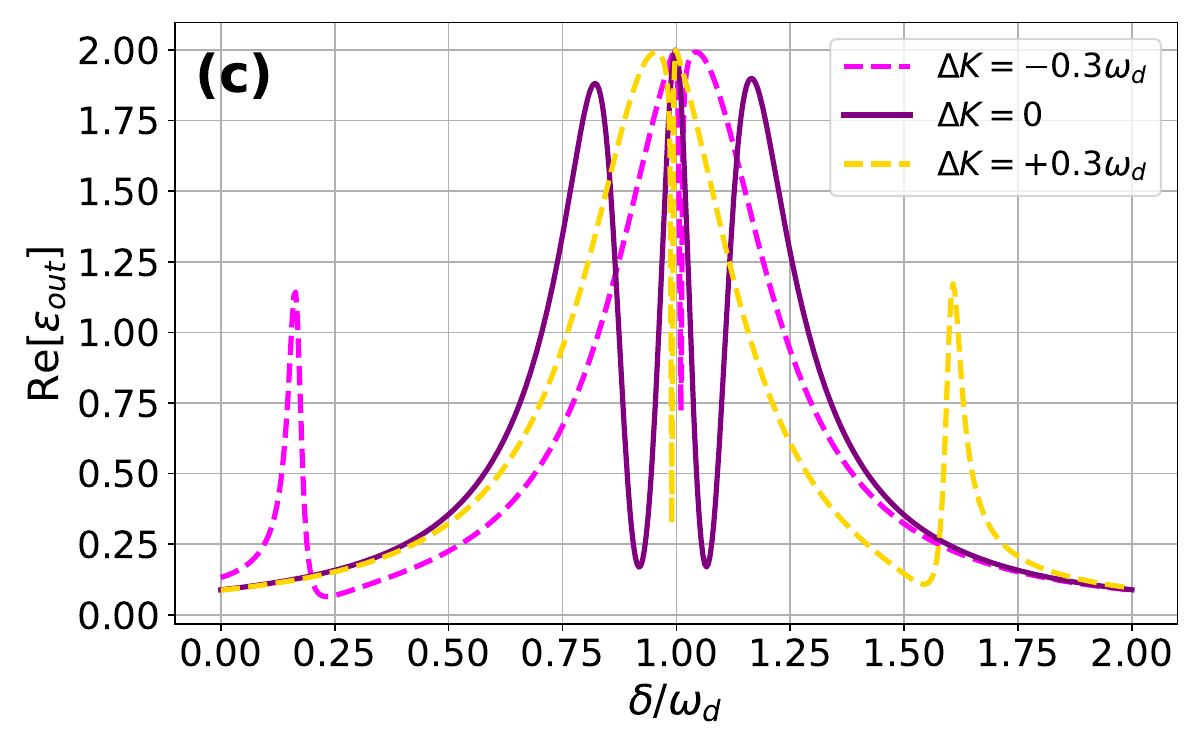}
			\includegraphics[scale=0.43]{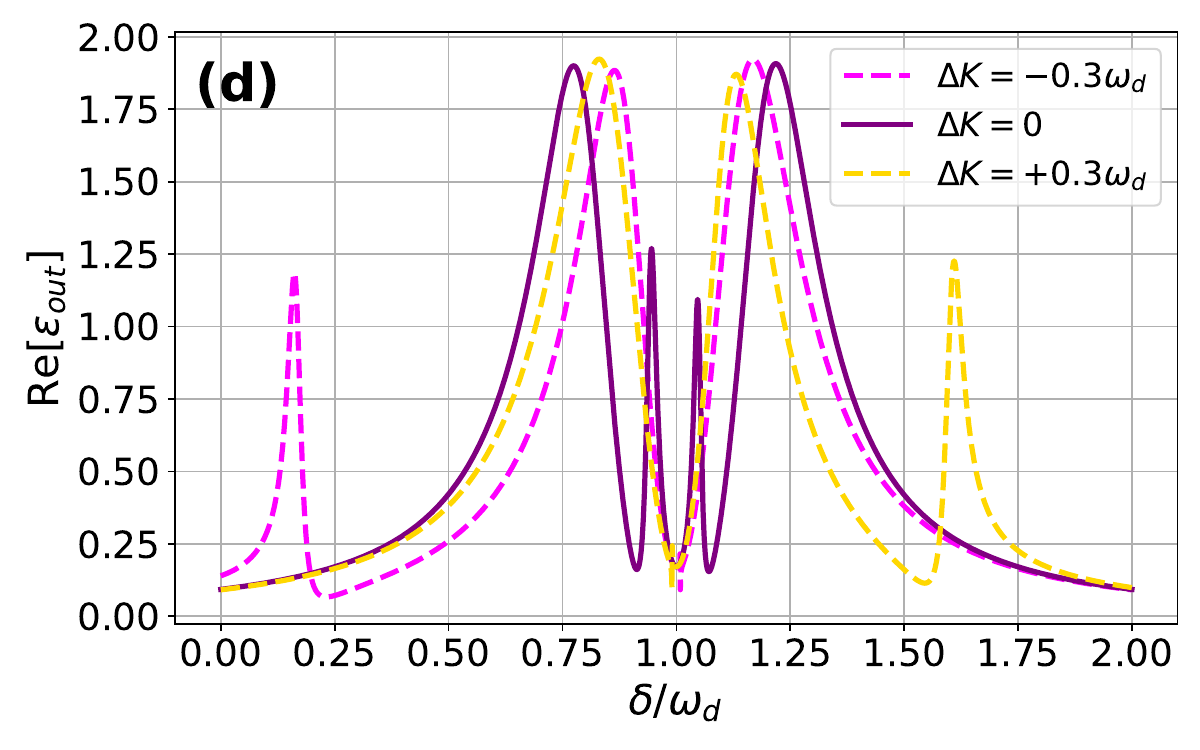}
			\includegraphics[scale=0.43]{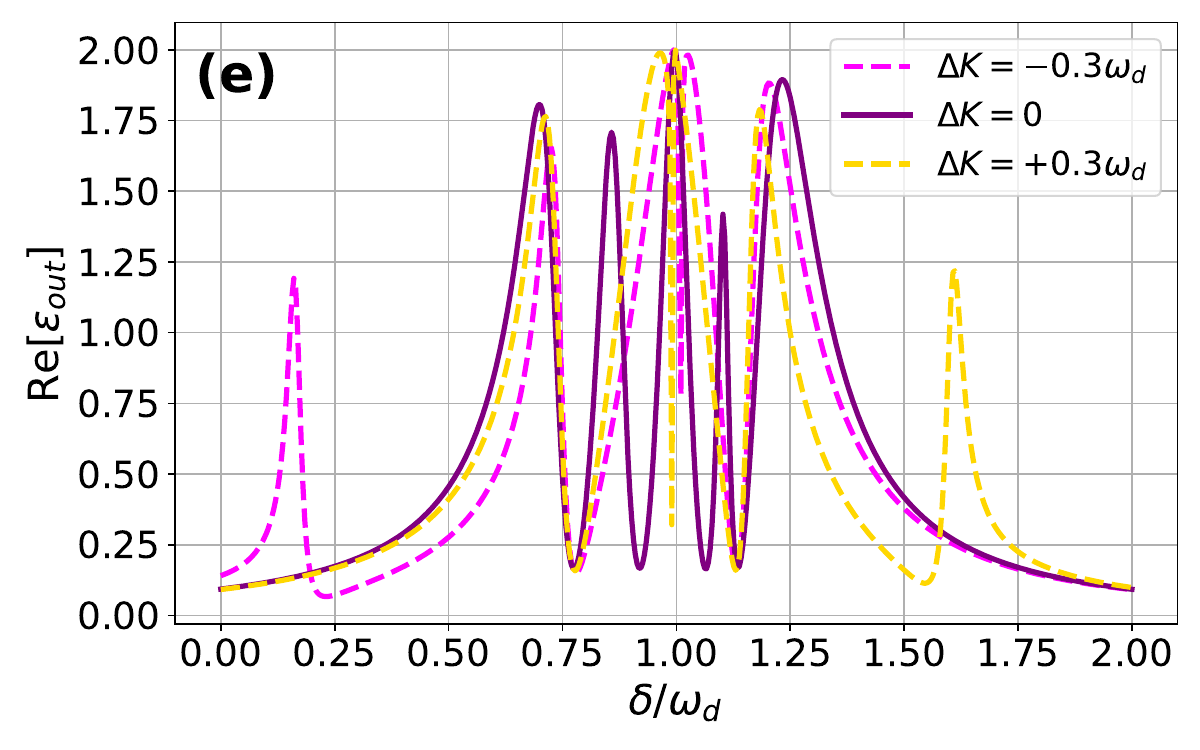}
			\includegraphics[scale=0.43]{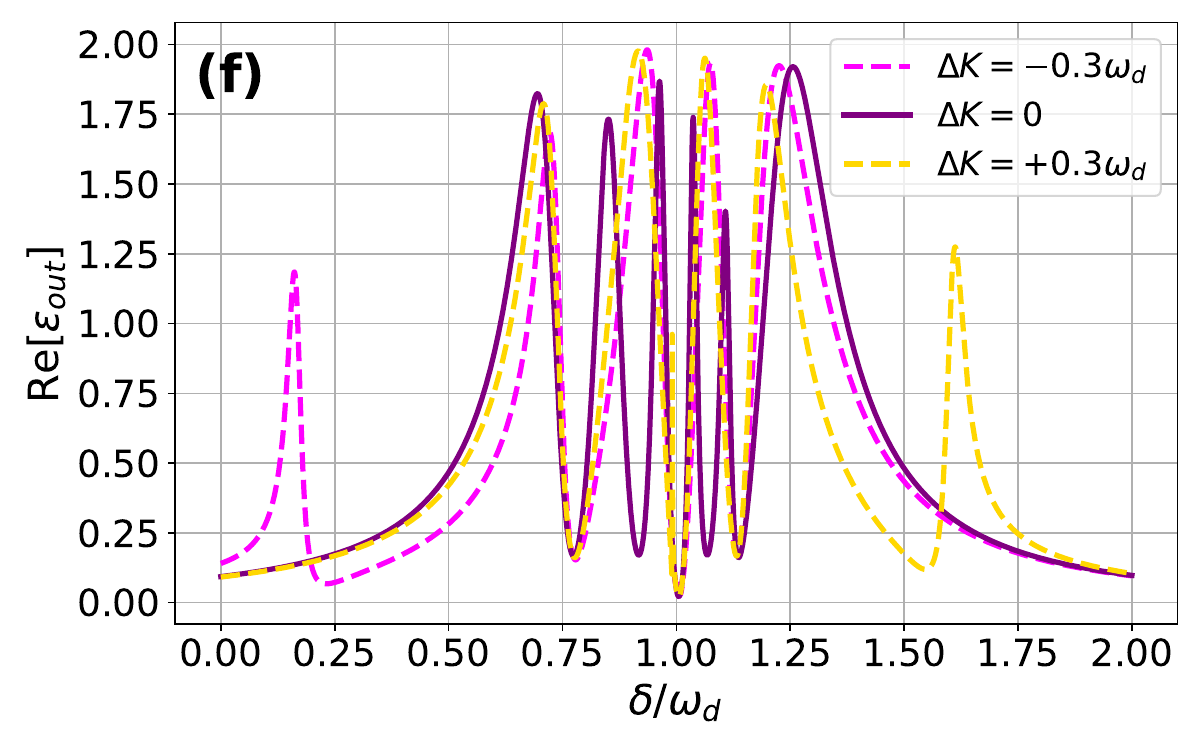}
			\caption{Real part of the output field $\epsilon_{\mathrm{R}}$ as a function of  normalized probe field frequency $\delta/\omega_{d}$. (a) $\chi_{1}=\chi_{2}=\zeta_{1}=\zeta_{2}=\zeta_a=0$; (b) $\chi_{1}/2\pi = 1.5$ MHz, $\chi_{2}=\zeta_{1}=\zeta_{2}=\zeta_a=0$; (c) $\chi_{1}/2\pi = \zeta_{1}/2\pi = 1.5$ MHz, $\chi_{2}=\zeta_{2}=\zeta_a=0$; (d) $\chi_{1}/2\pi = \chi_{2}/2\pi = \zeta_{1}/2\pi = 1.5$ MHz, $\zeta_{2}=\zeta_a=0$; (e) $\chi_{1}/2\pi = \chi_{2}/2\pi = \zeta_{1}/2\pi = 1.5$ MHz, $\zeta_{2}/2\pi = 3.5$ MHz, $\zeta_a=0$; (f) $\chi_{1}/2\pi = \chi_{2}/2\pi = \zeta_{2}/2\pi = 1.5$ MHz, $\zeta_{2}/2\pi = 3.5$ MHz, $\zeta_{a}/2\pi = 2.5$ MHz. All remaining parameters are given in the text.}\label{b}
		\end{center}
	\end{figure} 
	Figures~\ref{b}(a)–\ref{b}(f) display the absorption spectrum of the output probe field as a function of the normalized detuning \( \delta/\omega_d \) for different interaction configurations. To identify the physical origin of the transparency features, we first consider the case where the Kerr effect is absent ($\Delta K=0$), so that the spectral response is governed solely by coherent interactions between the photon, magnon, and phonon modes. When all couplings are switched off ($\chi_{1}=\chi_{2}=\zeta_{1}=\zeta_{2}=\zeta_a=0$), as shown in Figure~\ref{b}(a), the system behaves as an uncoupled cavity, and the probe field experiences simple absorption, leading to a single broad Lorentzian peak without any transparency window. In this case, the cavity mode is completely decoupled from the magnon and phonon modes, and therefore no destructive interference can occur. By activating only the photon-magnon coupling $\chi_{1}$, the absorption profile is significantly modified, as illustrated in Figure~\ref{b}(b), where a transparency dip emerges at the center of the spectrum, accompanied by two symmetric absorption maxima. This feature originates from destructive interference between the direct cavity excitation and the magnon-mediated excitation process, corresponding to magnon-induced transparency (MIT). When the magnon-phonon coupling $\zeta_{1}$ is further introduced, the first magnon mode becomes coupled to the corresponding mechanical mode through the magnetostrictive interaction. As a result, the single transparency window splits into two transparency windows, as observed in Figure~\ref{b}(c), which is characteristic of magnomechanically induced transparency (MMIT). This behavior originates from the coherent interaction between the magnon and phonon modes, which modifies the resonance condition and produces an additional destructive interference effect. The inclusion of the photon-magnon coupling $\chi_2$ further modifies the absorption spectrum, as shown in Figure~\ref{b}(d), where a third transparency window appears. Physically, the interaction between the cavity mode and the second magnon mode introduces an additional coherent interference process, leading to the emergence of this transparency window. Further introducing the magnon-phonon coupling $\zeta_{2}$ couples the second magnon mode to its corresponding mechanical mode. Consequently, four transparency windows appear in the absorption spectrum, as illustrated in Figure~\ref{b}(e). The simultaneous action of the two magnomechanical interactions further modifies the interference condition, leading to the emergence of an additional transparency window. Finally, incorporating the photon-phonon interaction $\zeta_a$ couples the cavity mode directly to the membrane vibration. As shown in Figure~\ref{b}(f), the absorption spectrum exhibits five distinct transparency windows. The central transparency window originates from the photon-phonon interaction and corresponds to optomechanically induced transparency, whereas the remaining transparency windows result from the combined cavity-magnon and magnon-phonon interactions. The coexistence of all coherent interactions enhances the interference responsible for the five-window transparency spectrum. We now turn to the influence of the Kerr effect ($\Delta K\neq 0$). In the absence of transparency windows, introducing the Kerr nonlinearity does not noticeably modify the absorption profile. However, once transparency windows are formed, the Kerr-induced frequency shift alters their spectral positions, where a positive Kerr shift moves the right transparency window toward higher detuning values, while a negative Kerr shift shifts the left transparency window toward lower detuning values. Consequently, the transparency spectrum becomes asymmetric, indicating the emergence of Fano-type resonances due to the Kerr-induced modification of the interference condition.\\
	\begin{figure}
		\begin{center}
			\includegraphics[scale=0.43]{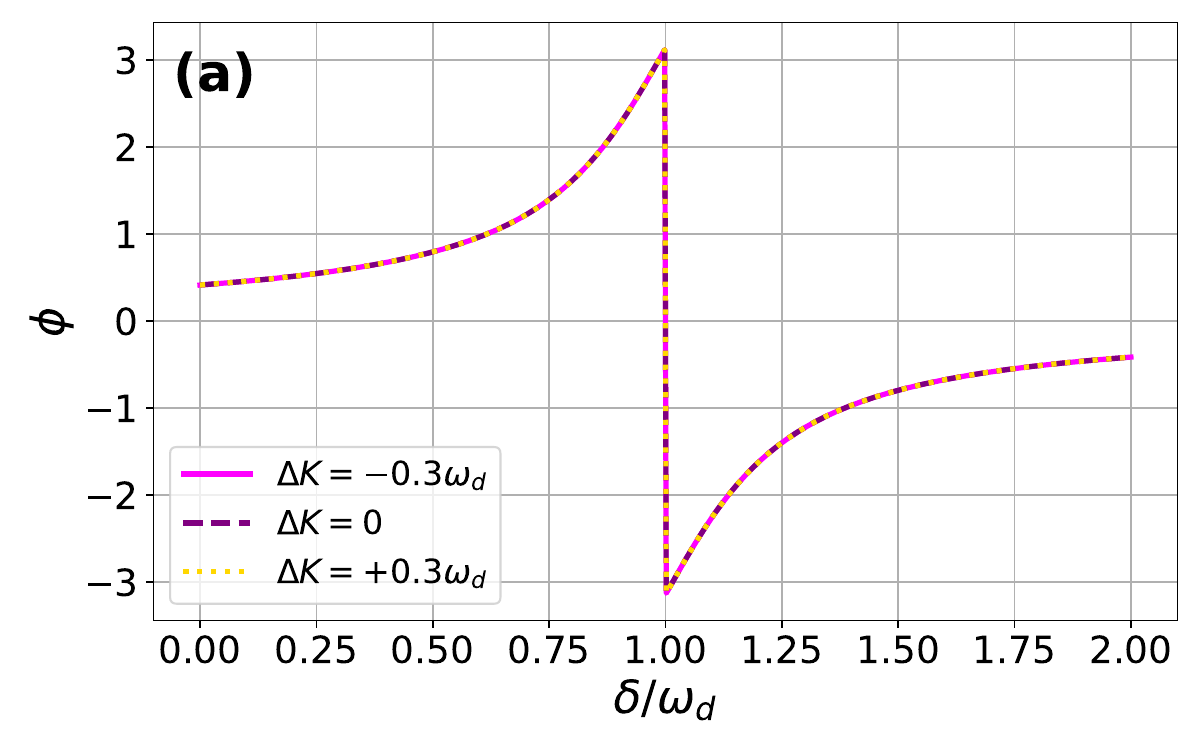}
			\includegraphics[scale=0.43]{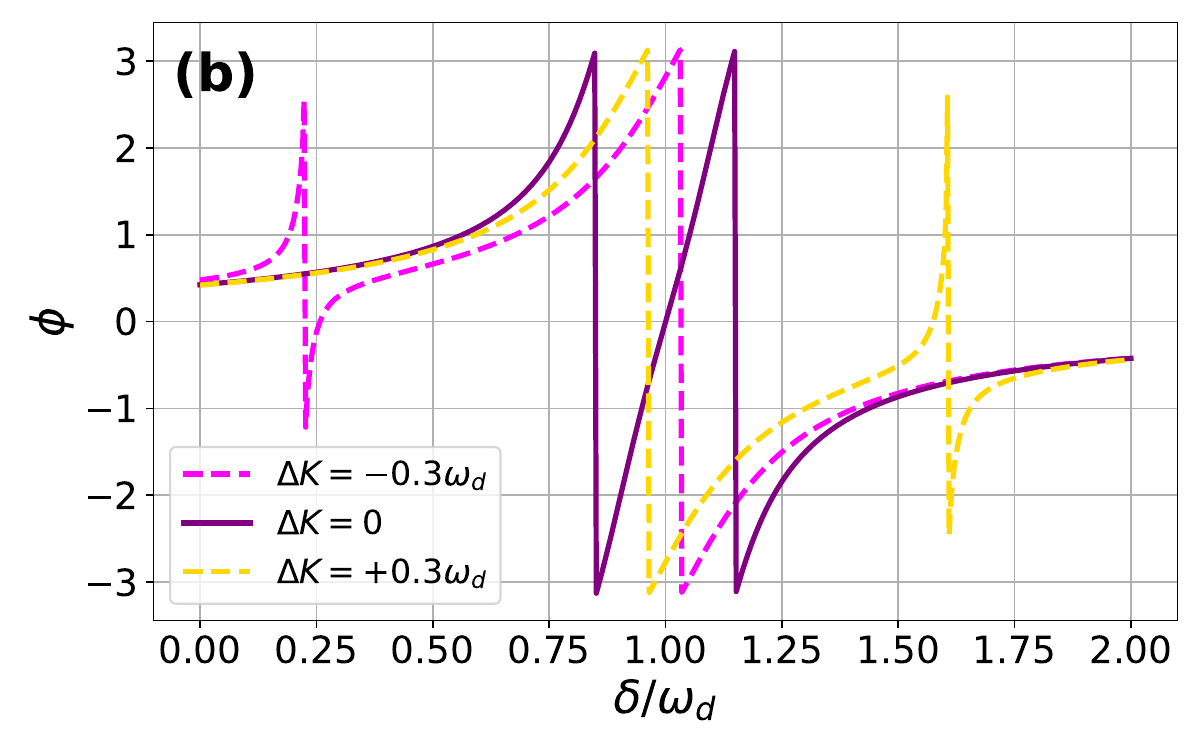}
			\includegraphics[scale=0.43]{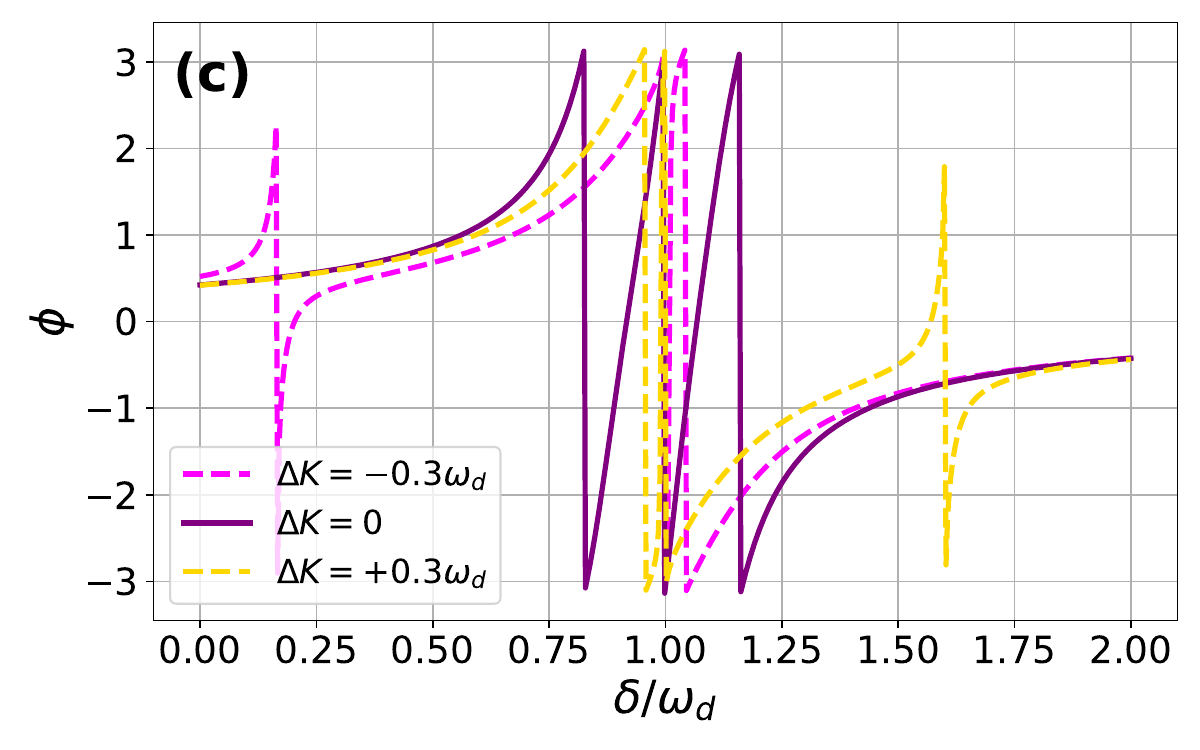}
			\includegraphics[scale=0.43]{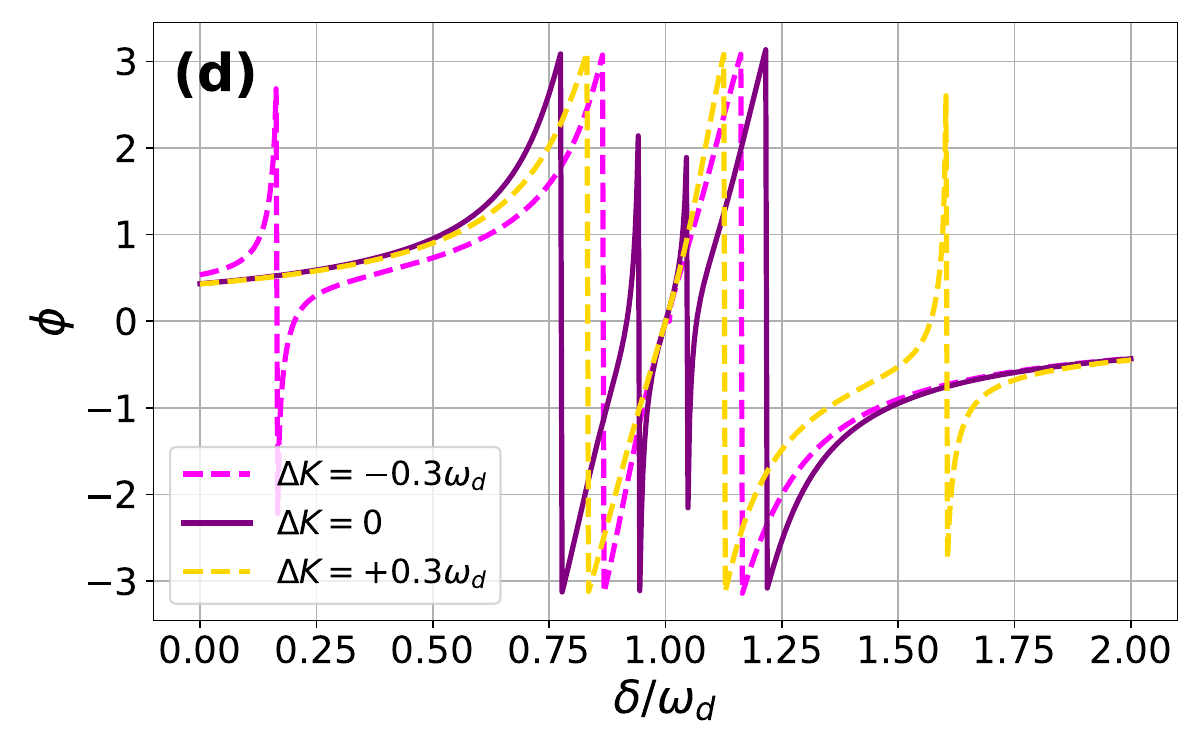}
			\includegraphics[scale=0.43]{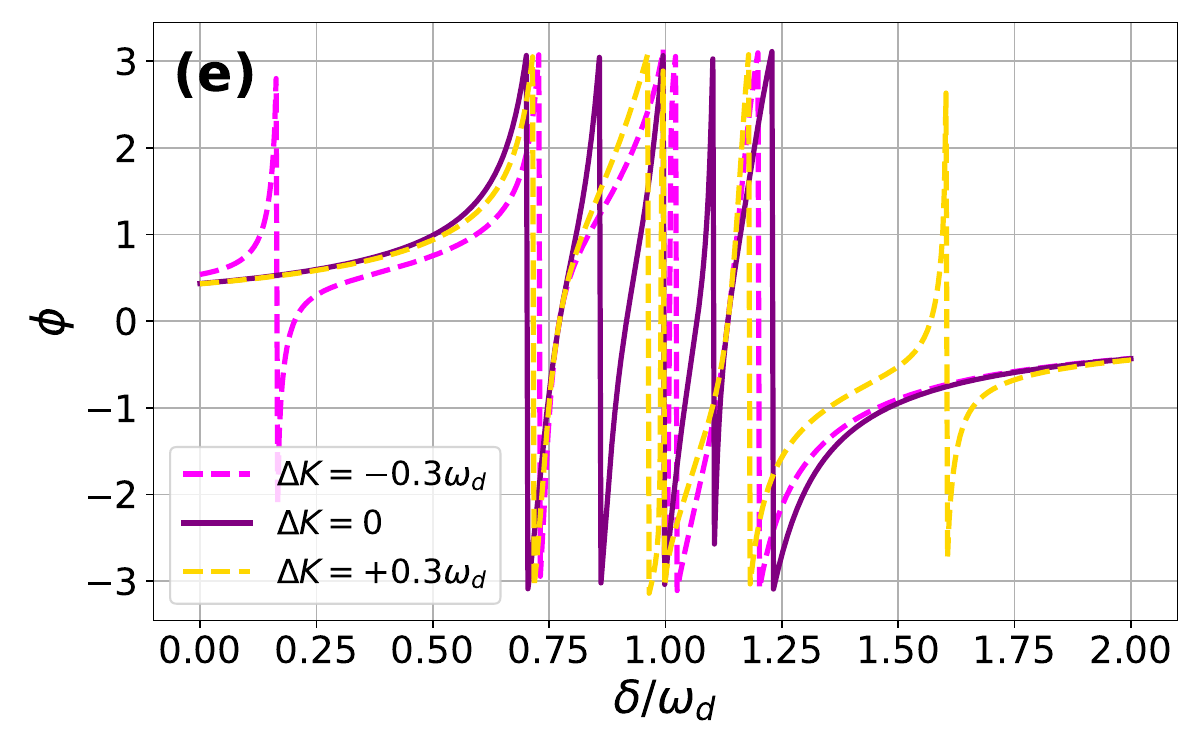}
			\includegraphics[scale=0.43]{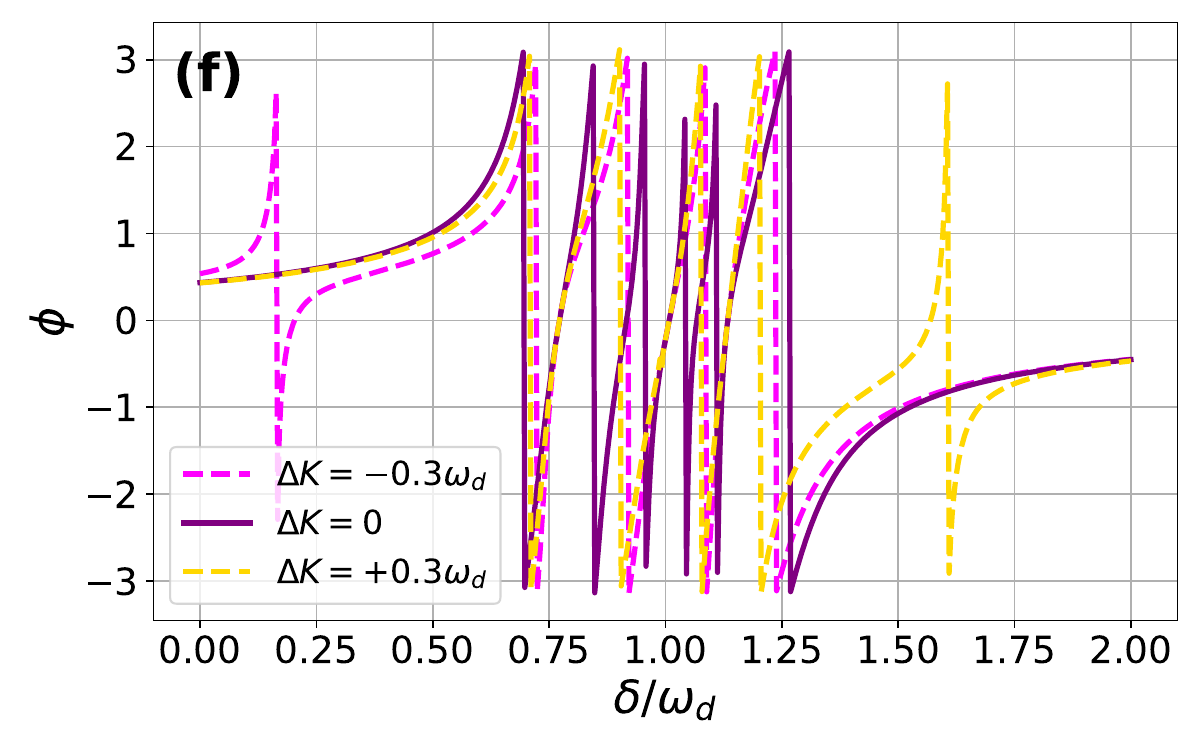}
			\caption{The phase $\phi$ of the transmitted probe field as a function of  normalized probe field frequency $\delta/\omega_{d}$. (a) $\chi_{1}=\chi_{2}=\zeta_{1}=\zeta_{2}=\zeta_a=0$; (b) $\chi_{1}/2\pi = 1.5$ MHz, $\chi_{2}=\zeta_{1}=\zeta_{2}=\zeta_a=0$; (c) $\chi_{1}/2\pi = \zeta_{1}/2\pi = 1.5$ MHz, $\chi_{2}=\zeta_{2}=\zeta_a=0$; (d) $\chi_{1}/2\pi = \chi_{2}/2\pi = \zeta_{1}/2\pi = 1.5$ MHz, $\zeta_{2}=\zeta_a=0$; (e) $\chi_{1}/2\pi = \chi_{2}/2\pi = \zeta_{1}/2\pi = 1.5$ MHz, $\zeta_{2}/2\pi = 3.5$ MHz, $\zeta_a=0$; (f) $\chi_{1}/2\pi = \chi_{2}/2\pi = \zeta_{2}/2\pi = 1.5$ MHz, $\zeta_{2}/2\pi = 3.5$ MHz, $\zeta_{a}/2\pi = 2.5$ MHz. All remaining parameters are given in the text.}\label{d}
		\end{center}
	\end{figure} 
	Figures~\eqref{d}(a)–\eqref{d}(f) present the phase \( \phi \) of the transmitted probe field as a function of the normalized detuning \( \delta/\omega_d \) for different coupling configurations and Kerr frequency shifts. In Figure~\eqref{d}(a), all interaction parameters are set to zero, and the phase exhibits a single dispersive profile with a sharp phase jump around resonance, which corresponds to the absence of induced transparency. When the photon–magnon coupling \( \chi_1 \) is activated while the remaining couplings are kept at zero, as shown in Figure~\eqref{d}(b), the phase spectrum develops an additional rapid variation, indicating the emergence of a single transparency window associated with destructive interference between the cavity mode and the first magnon mode. With the inclusion of the magnon-phonon coupling \( \zeta_1 \), shown in Figure~\eqref{d}(c), the phase response evolves into two pronounced dispersive features, reflecting the emergence of a second transparency window. This behavior originates from the coherent interaction between the first magnon mode and its corresponding mechanical mode via the magnetostrictive effect. As shown in Figure~\eqref{d}(d), the simultaneous presence of the photon-magnon \( \chi_1 \) and magnon-phonon \( \chi_2 \) couplings further reshapes the phase response, leading to three pronounced phase variations. The additional phase transition reflects the emergence of a third transparency window resulting from the coherent interactions among the cavity, magnon, and phonon modes. The introduction of the second magnon-phonon coupling \( \zeta_2 \), as depicted in Figure~\eqref{d}(e), further modifies the phase response, giving rise to four pronounced phase variations. This behavior originates from the coherent interaction between the second magnon mode and its corresponding mechanical mode, leading to the emergence of a fourth transparency window. Finally, when all couplings, including the photon-phonon interaction \( \zeta_a \), are activated, the phase response shown in Figure~\eqref{d}(f) exhibits five pronounced dispersive features corresponding to the five transparency windows. This behavior results from the combined photon-magnon, magnon-phonon, and photon-phonon interactions. Moreover, the effect of the Kerr shift $\Delta K$ is clearly visible in all panels. A positive Kerr shift moves the phase transitions toward higher detuning values, whereas a negative Kerr shift shifts them toward lower detuning values. As a result, the phase response becomes asymmetric. The steep phase variations observed around the transparency windows indicate a pronounced dispersive response, whose spectral position is controlled by the Kerr-induced frequency shift.\\
	\begin{figure}
		\begin{center}
			\includegraphics[scale=0.51]{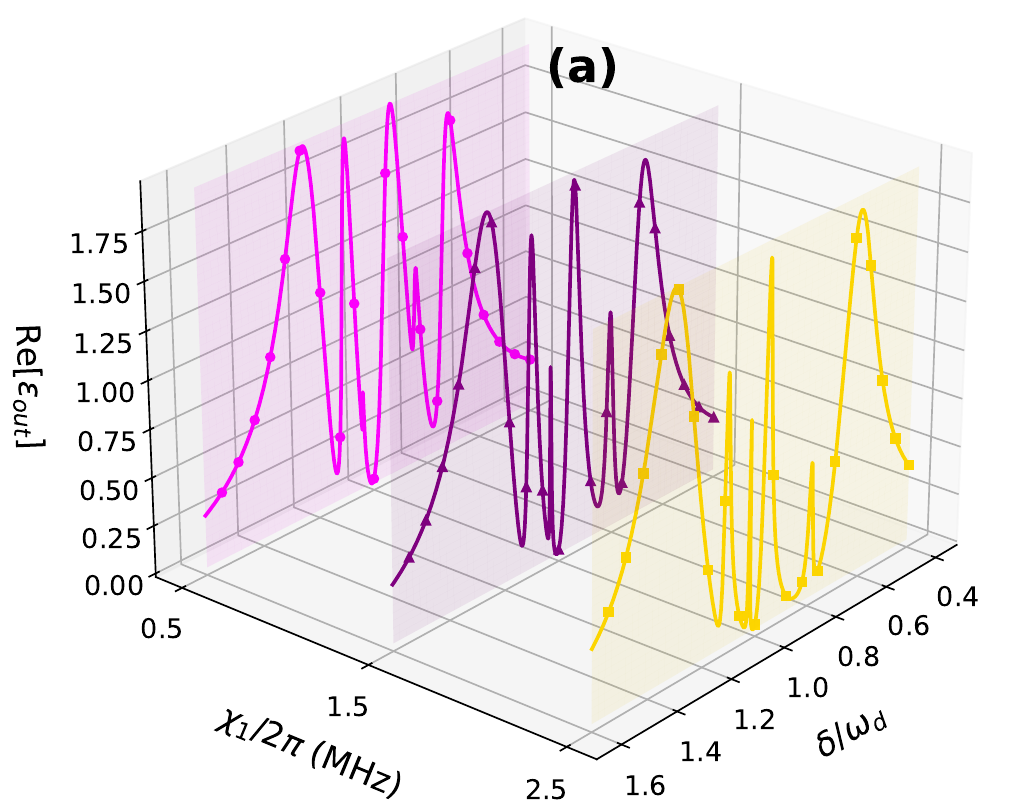}
			\includegraphics[scale=0.51]{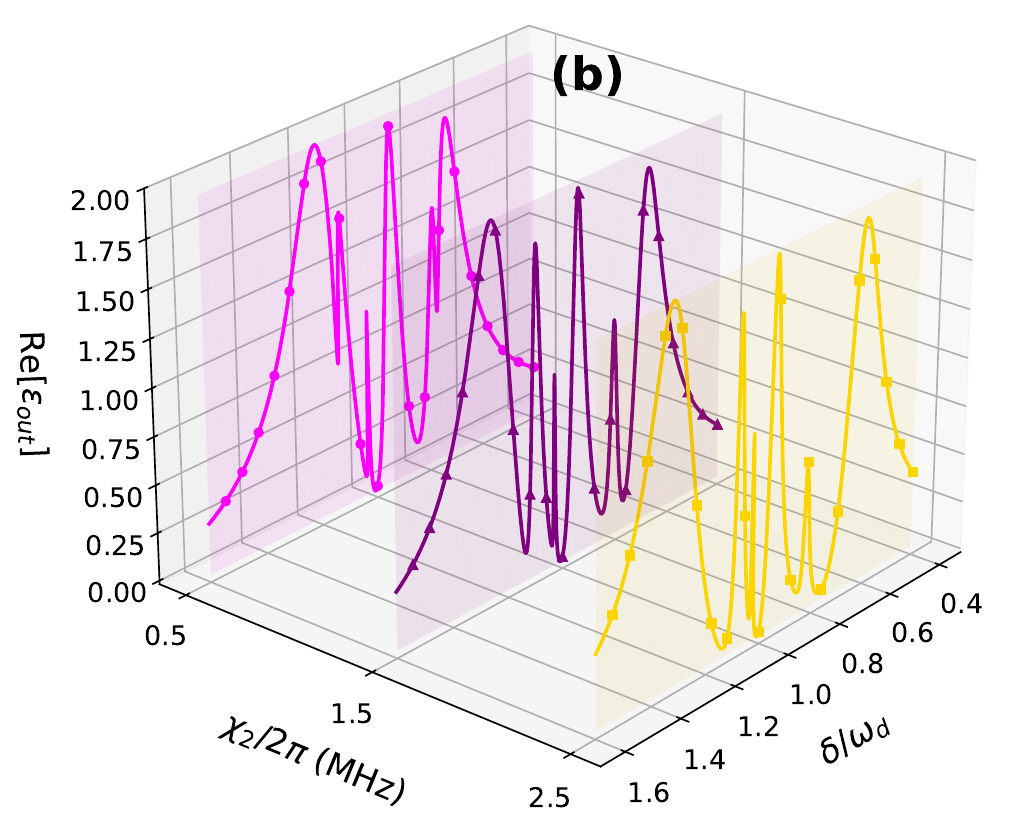}
			\includegraphics[scale=0.50]{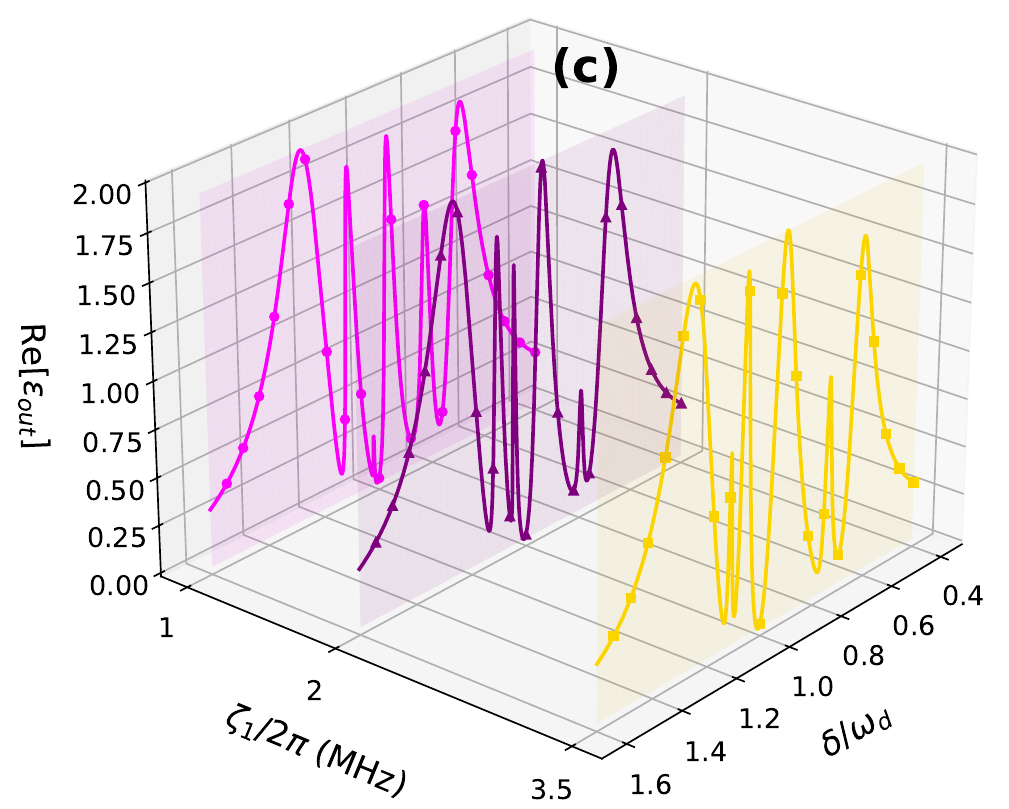}
			\includegraphics[scale=0.51]{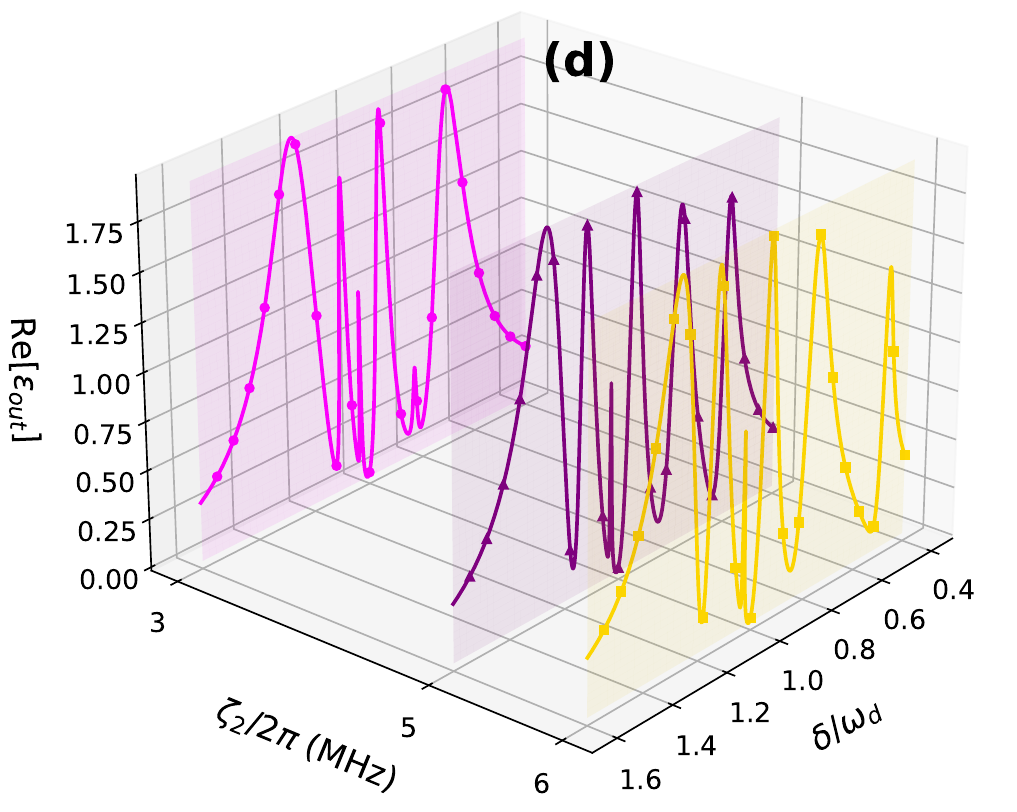}
			\includegraphics[scale=0.51]{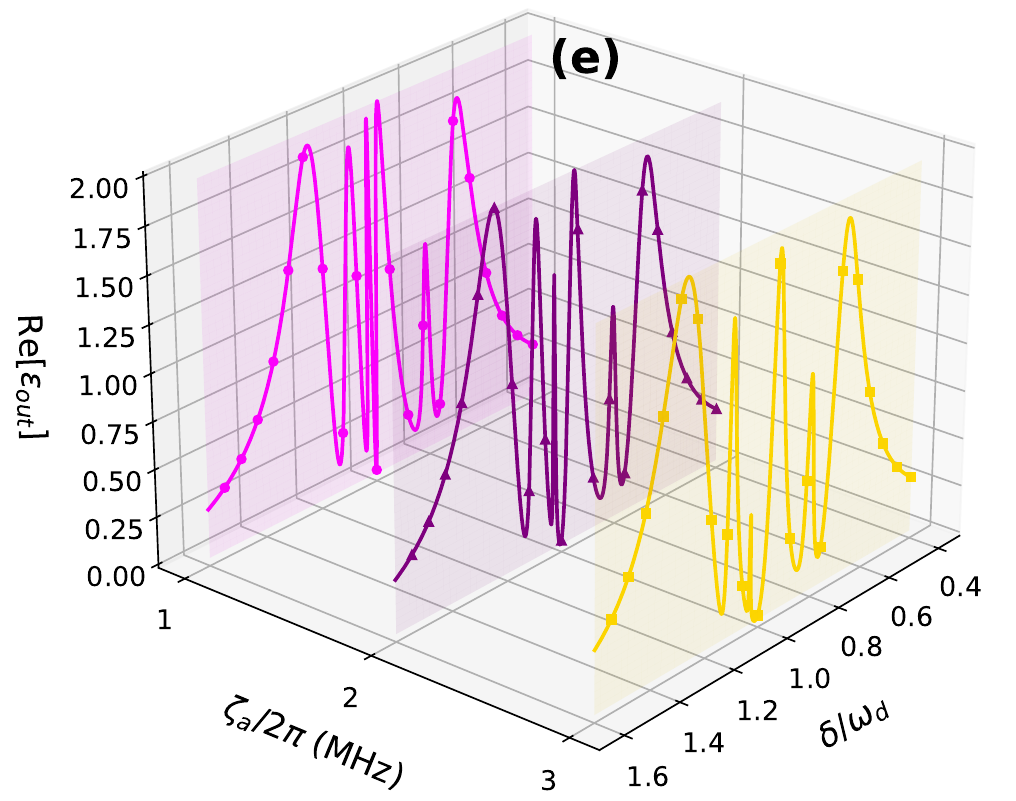}
			\caption{Absorption of the output field \(\text{Re}[\epsilon_{\mathrm{out}}]\) plotted versus the normalized detuning $\delta/\omega_d$ for different interaction strengths, with $\Delta K/2\pi = -0.5$ MHz. The remaining parameters are listed in the text.}\label{dd}
		\end{center}
	\end{figure} 
	Figures~\ref{dd}(a)–\ref{dd}(e) show the absorption profile $\mathrm{Re}[\epsilon_{out}]$ of the output probe field as a function of the normalized detuning $\delta/\omega_d$ for different values of the coupling strengths, with $\Delta K/2\pi=-0.5$ MHz. In Figure~\ref{dd}(a), the magnon-photon coupling $\chi_1$ is varied, whereas the other parameters are kept fixed at $\zeta_1/2\pi=\chi_2/2\pi=1.5$ MHz, $\zeta_2/2\pi=3.5$ MHz, and $\zeta_a/2\pi=2.5$ MHz. As $\chi_1/2\pi$ is increased from $0.5$ MHz to $2.5$ MHz, the second and fourth transparency windows become progressively broader and deeper, indicating that the interaction between the cavity mode and the first magnon mode is strengthened. Consequently, the destructive interference responsible for these transparency windows becomes more pronounced. In Figure~\ref{dd}(b), the second magnon-photon coupling $\chi_2$ is varied under the same parameter conditions. As $\chi_2/2\pi$ is increased from $0.5$ MHz to $2.5$ MHz, the first and fifth transparency windows become progressively broader and deeper, whereas the central three transparency windows remain nearly unchanged. This evolution indicates that a stronger $\chi_2$ reinforces the destructive interference between the cavity mode and the second magnon mode. Figure~\ref{dd}(c) illustrates the effect of the magnon-phonon coupling $\zeta_1$, while $\chi_1/2\pi=\chi_2/2\pi=1.5$ MHz, $\zeta_2/2\pi=3.5$ MHz, and $\zeta_a/2\pi=2.5$ MHz are kept fixed. Five distinct transparency windows are clearly observed at well-defined values of the normalized detuning $\delta/\omega_d$. As $\zeta_1/2\pi$ is increased from $1.5$ MHz to $3.5$ MHz, the second transparency window shifts toward lower detuning, whereas the fourth shifts toward higher detuning, while the remaining three remain nearly unchanged. This behavior indicates that the magnon-phonon coupling $\zeta_1$ governs the spectral positions of the second and fourth transparency windows. The influence of the second magnon-phonon coupling $\zeta_2$ is presented in Figure~\ref{dd}(d). As in the previous case, varying the magnon-phonon coupling modifies the spectral positions of the corresponding transparency windows.Specifically, as $\zeta_2/2\pi$ is increased from $1.5$ MHz to $3.5$ MHz, the first transparency window shifts toward lower detuning, whereas the fifth shifts toward higher detuning, while the remaining three remain nearly unchanged. This behavior demonstrates that their spectral positions are governed by the magnon-phonon coupling $\zeta_2$.
    Finally, Figure~\ref{dd}(e) demonstrates the role of the photon-phonon coupling $\zeta_a$, while $\chi_1/2\pi=\chi_2/2\pi=\zeta_1/2\pi=1.5$ MHz and $\zeta_2/2\pi=3.5$ MHz are kept fixed. As $\zeta_a/2\pi$ is increased from $1$ MHz to $3$ MHz, the central transparency windows become progressively broader and deeper, whereas the outer ones remain nearly unchanged. The enhancement of the central transparency windows indicates that the photon-phonon interaction reinforces the destructive interference associated with the optomechanical interaction.\\
	\begin{figure}
		\begin{center}
			\includegraphics[scale=0.43]{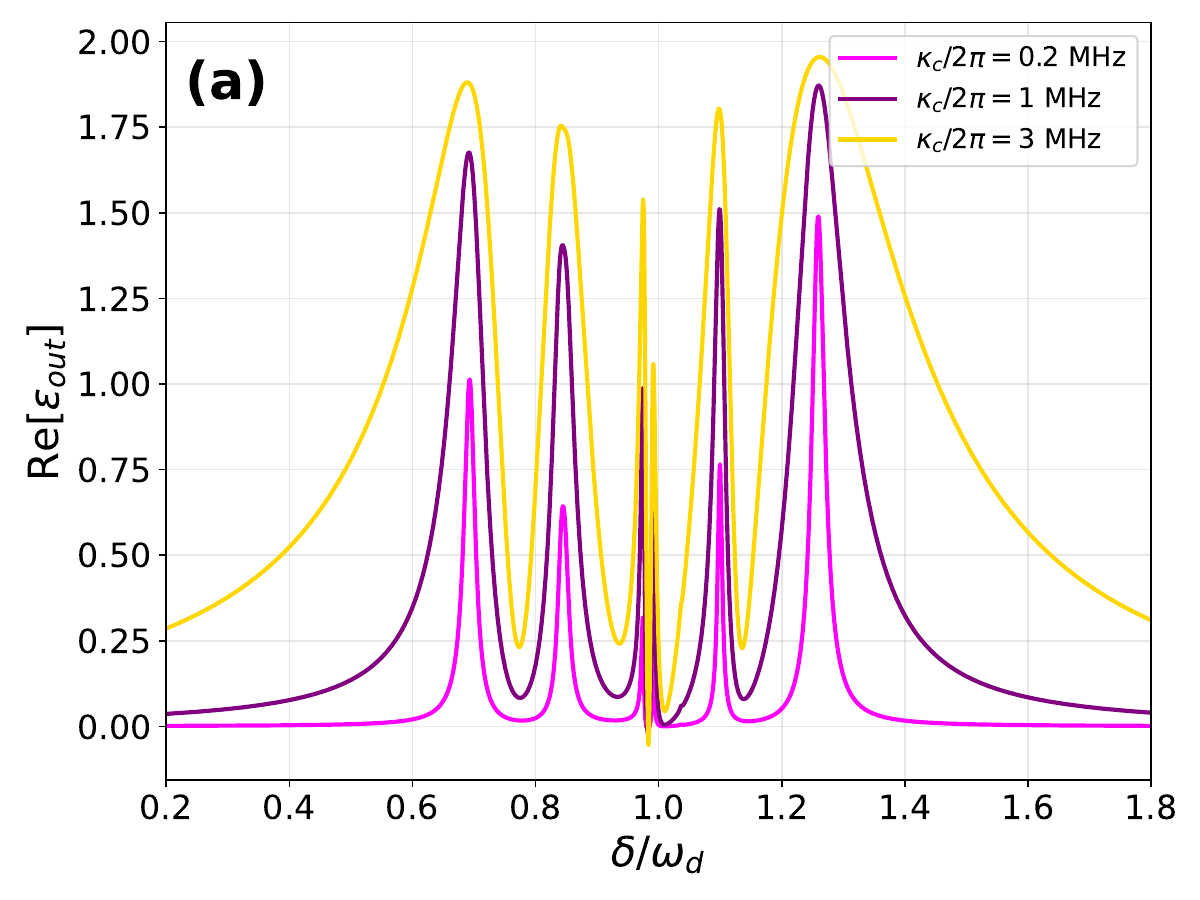}
			\includegraphics[scale=0.43]{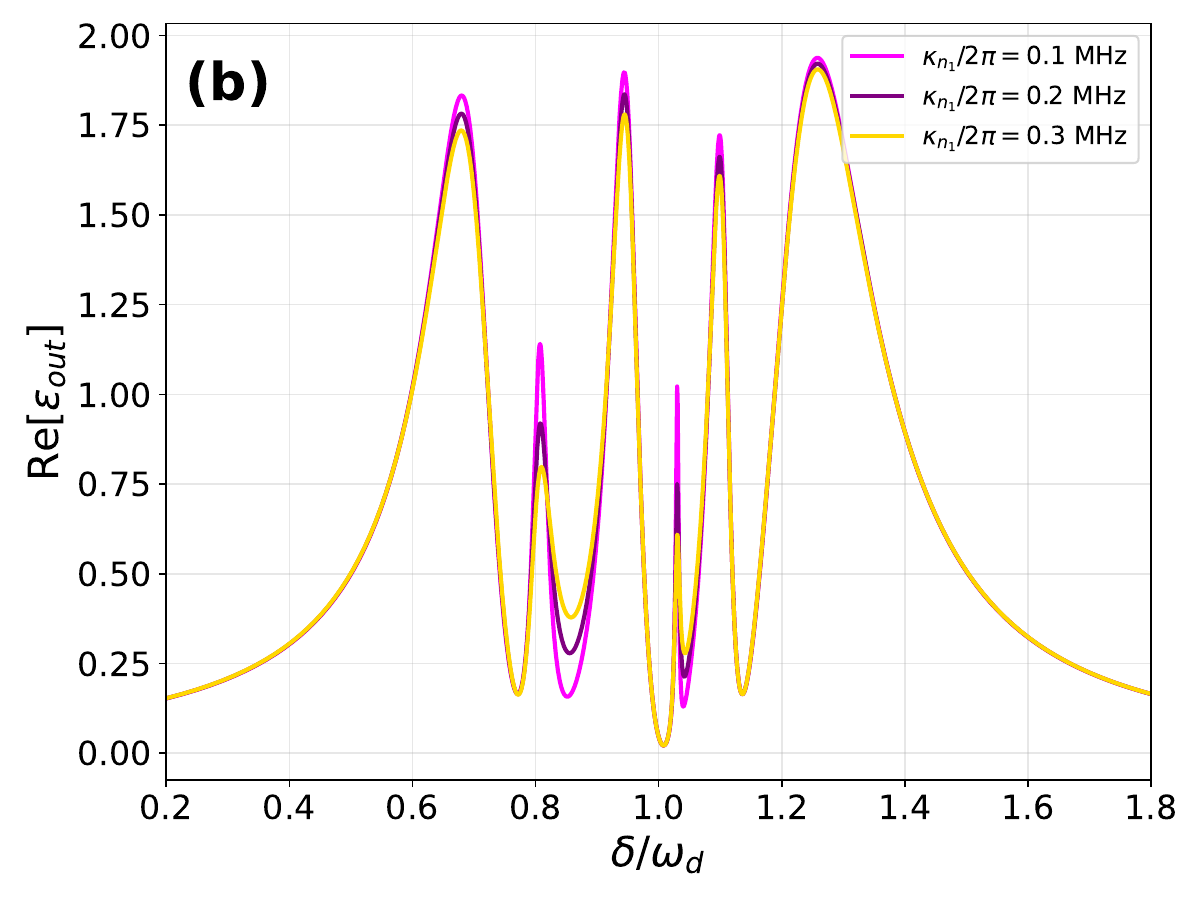}
			\includegraphics[scale=0.43]{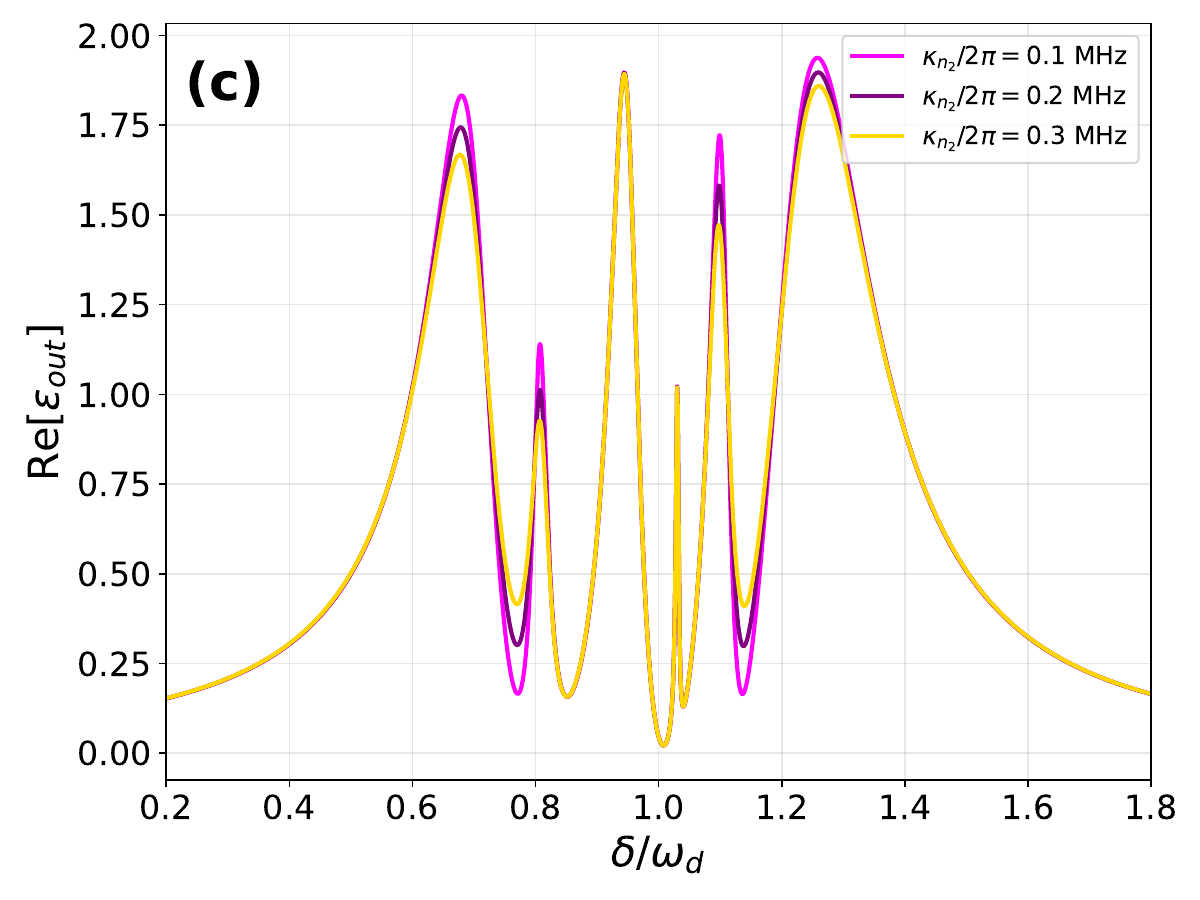}
			\caption{Absorption of the output field $\mathrm{Re}[\epsilon_{\mathrm{out}}]$ as a function of the normalized detuning $\delta/\omega_d$ for different dissipation parameters: (a) varying the cavity decay rate, (b) varying the damping rate of the first magnon mode $n_1$, and (c) varying the damping rate of the second magnon mode $n_2$. The remaining parameters are given in the text.}\label{ddd0}
		\end{center}
	\end{figure}
	Figures~\ref{ddd0}(a)--\ref{ddd0}(c) show the variation of the absorption spectrum $\varepsilon_R$ of the output probe field for different decay rates and dissipation parameters as a function of the normalized probe detuning $\delta/\omega_d$. Figure~\ref{ddd0}(a) presents the absorption profile for several values of the cavity decay rate, while the remaining parameters are fixed as specified in the main text. It is evident that reducing the cavity decay rate leads to broader and deeper transparency windows. This behavior can be attributed to the longer photon lifetime inside the cavity at lower cavity decay rates. The reduced cavity losses preserve the coherence of the intracavity field, thereby strengthening the destructive interference responsible for the transparency effect and producing more pronounced transparency windows. Figures~\ref{ddd0}(b) and~\ref{ddd0}(c) illustrate the effect of the magnon dissipation rates on the absorption spectrum. In Figure~\ref{ddd0}(b), decreasing the dissipation rate of the first magnon mode $n_1$ significantly enhances the second and fourth transparency windows, making them broader and deeper. This behavior indicates that these two transparency windows are primarily controlled by the interaction between the cavity mode and the first magnon mode. Similarly, Figure~\ref{ddd0}(c) shows that reducing the dissipation rate of the second magnon mode $n_2$ enhances the first and fifth transparency windows, whereas the remaining three are only weakly affected. The enhancement of these two transparency windows indicates that lowering the dissipation rate of the second magnon mode strengthens the destructive interference associated with the cavity-second-magnon interaction. Overall, decreasing the magnon dissipation rates enhances the effective photon-magnon interaction, thereby reinforcing the destructive interference responsible for the transparency effect and yielding more pronounced transparency windows.
	\subsection{The fast and slow light effects } \label{xxx}
In this subsection, we analyze the influence of the photon-magnon couplings ($\chi_1$, $\chi_2$), the magnon-phonon couplings ($\zeta_1$, $\zeta_2$), the photon-phonon coupling ($\zeta_a$), and the Kerr-induced frequency shift ($\Delta K$) on the group delay of the output probe field. The phase of the transmitted probe field, defined in Eq.~\eqref{T}, is directly related to the group delay $\tau$ of the output field through
	\begin{equation}\label{cv}
		\tau = \frac{\partial \phi}{\partial \omega_p}
		= \mathrm{Im}\!\left[ T^{-1} \frac{\partial T}{\partial \omega_p} \right],
	\end{equation}
	A negative phase slope corresponds to a negative group delay, characterizing the fast light effect ($\tau_g<0$). In contrast, a positive phase of the transmitted phase leads to a positive group delay, which is associated with slow light propagation ($\tau_g>0$).\\
	\begin{figure}
		\begin{center}
			\includegraphics[scale=0.43]{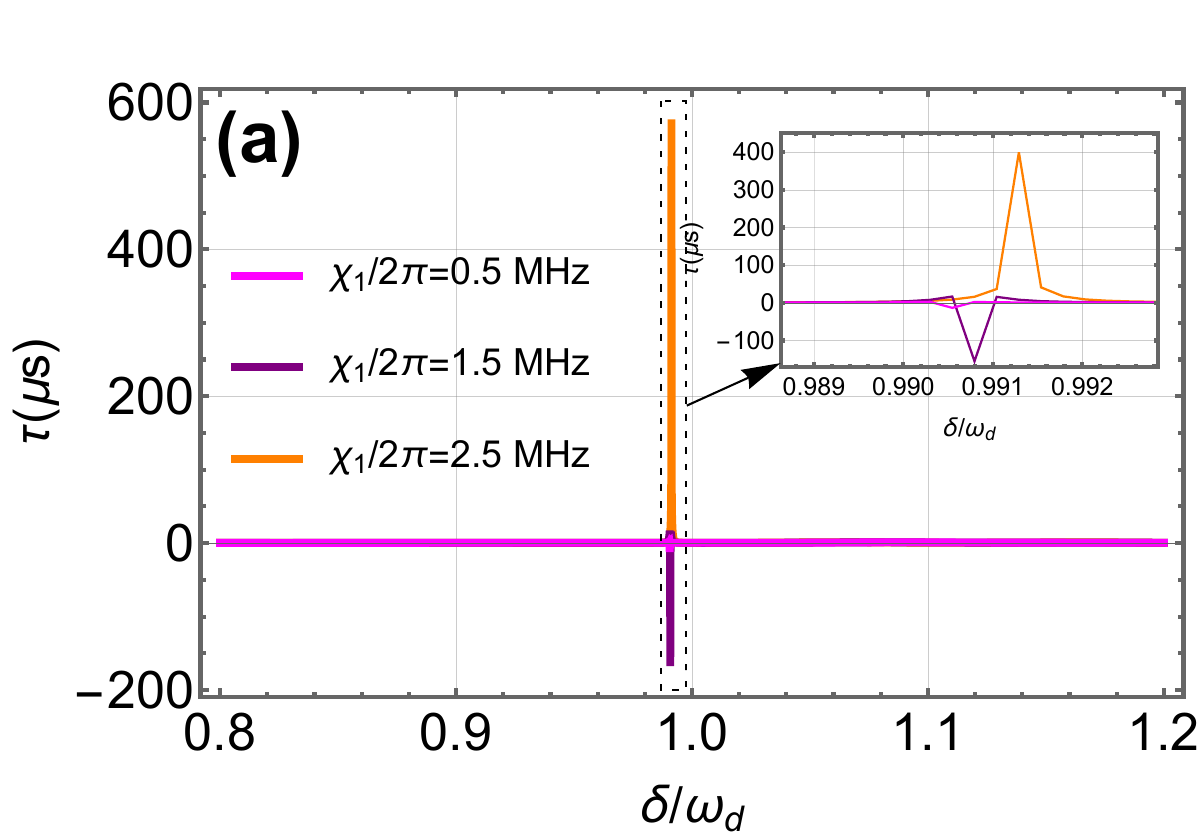}
			\includegraphics[scale=0.43]{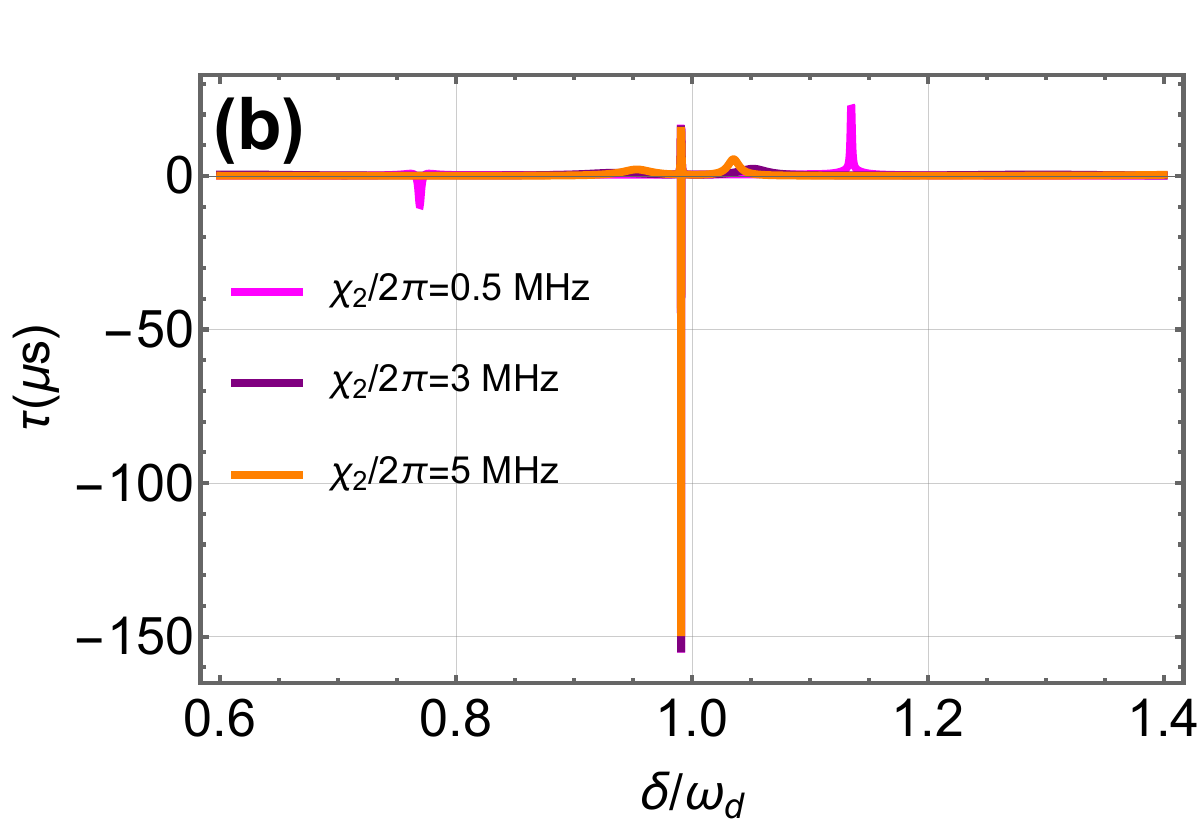}
			\includegraphics[scale=0.43]{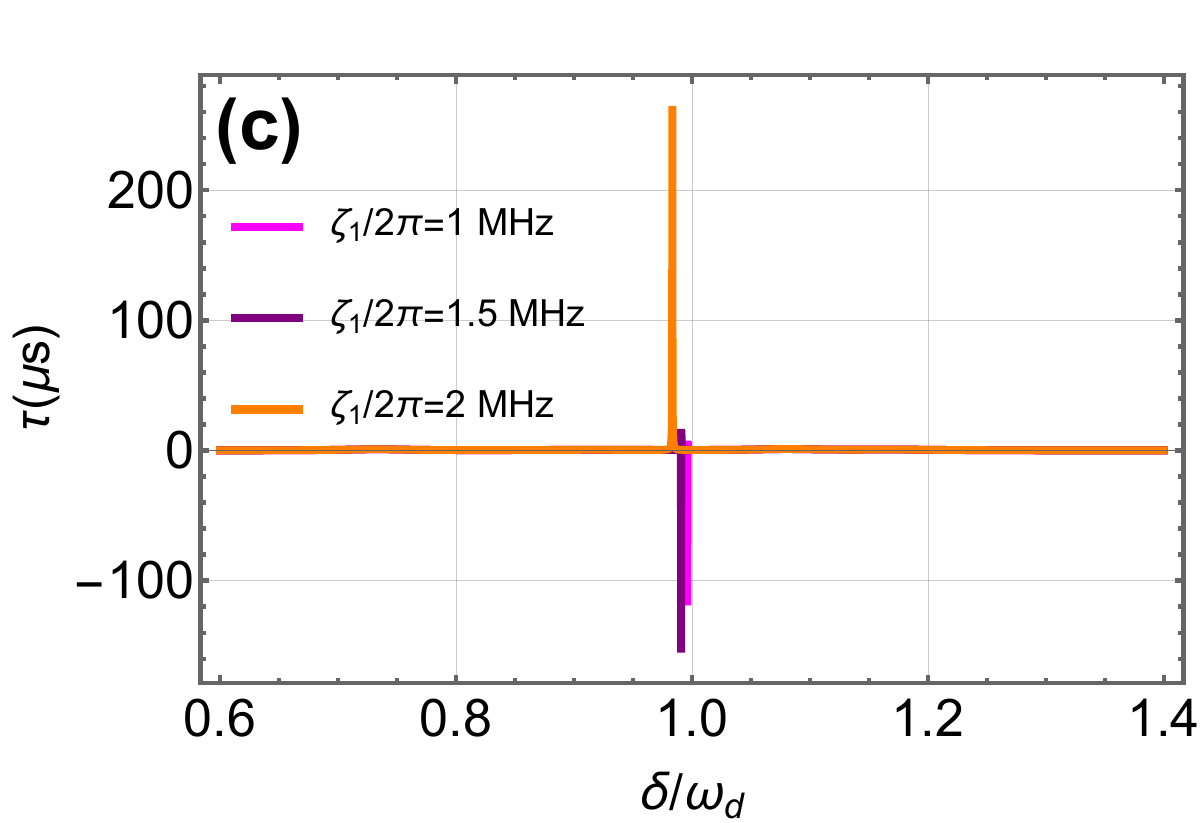}
			\includegraphics[scale=0.43]{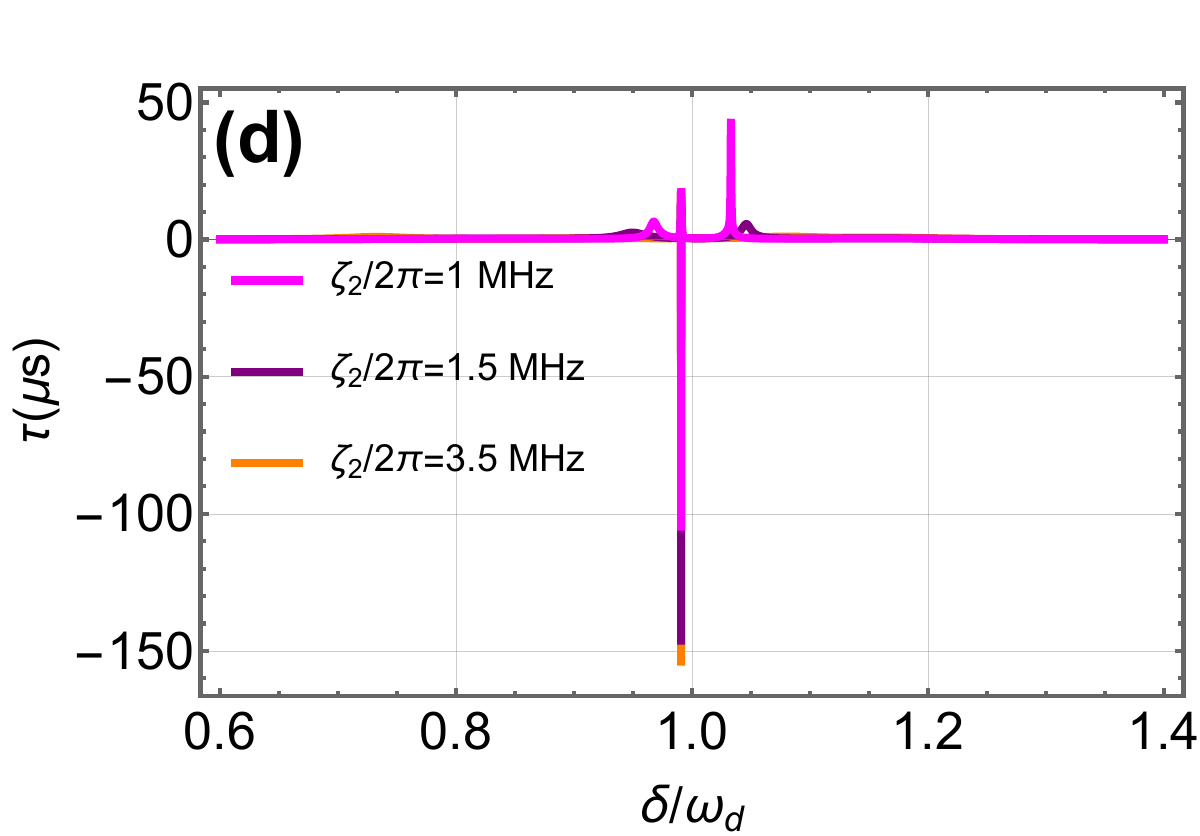}
			\includegraphics[scale=0.44]{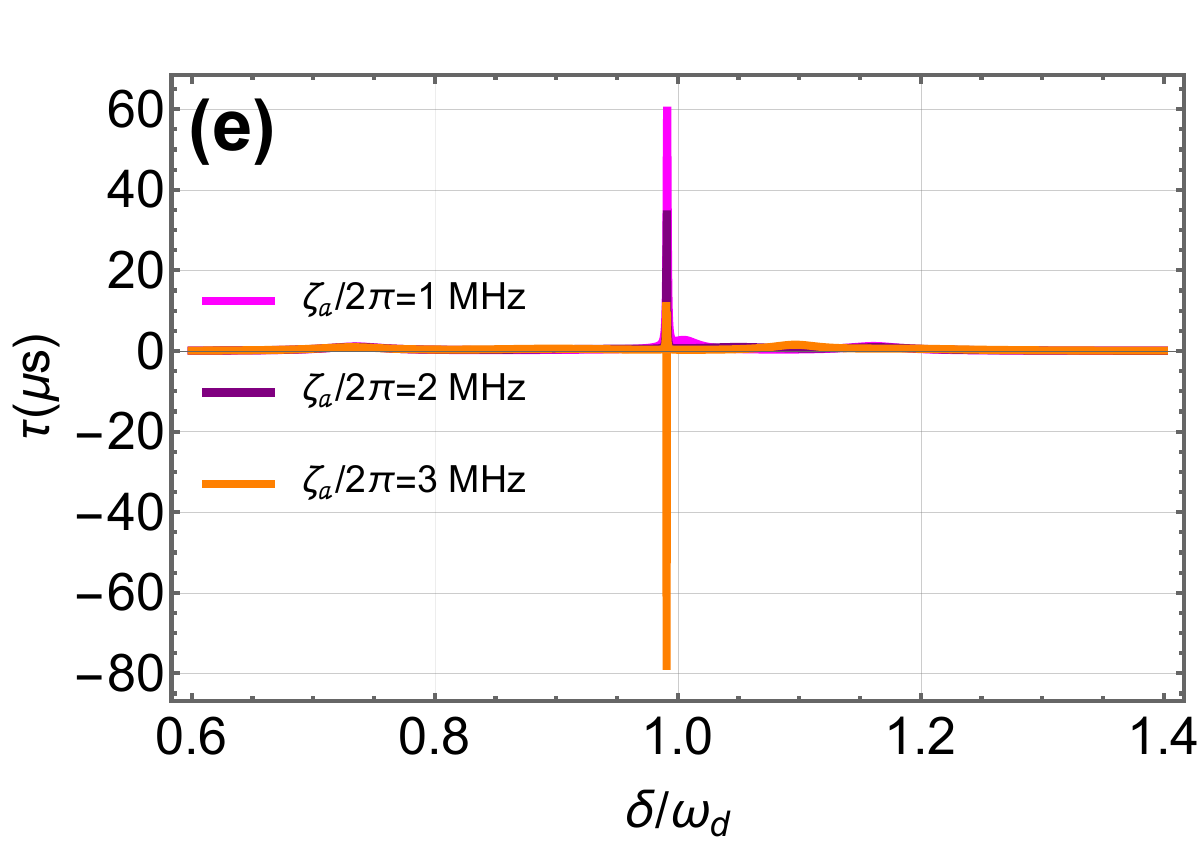}
			\caption{Group delay $\tau$ of the transmitted probe field plotted versus the normalized detuning $\delta/\omega_d$ for several coupling strengths, with $\Delta K = 0.3\,\omega_d$. (a) varying $\chi_1$ while $\chi_2/2\pi = \zeta_1/2\pi = 1.5$ MHz, $\zeta_2/2\pi = 3.5$ MHz, and $\zeta_a/2\pi = 2.5$ MHz are kept constant. (b) varying $\chi_2$ with $\chi_1/2\pi = \zeta_1/2\pi = 1.5$ MHz, $\zeta_2/2\pi = 3.5$ MHz, and $\zeta_a/2\pi = 2.5$ MHz. (c) varying $\zeta_1$ values with $\chi_1/2\pi = \chi_2/2\pi = 1.5$ MHz, $\zeta_2/2\pi = 3.5$ MHz, and $\zeta_a/2\pi = 2.5$ MHz. (d) varying $\zeta_2$ values when $\chi_1/2\pi = \chi_2/2\pi = \zeta_1/2\pi = 1.5$ MHz and $\zeta_a/2\pi = 2.5$ MHz. (e) varying $\zeta_a$ with $\chi_1/2\pi = \chi_2/2\pi = \zeta_1/2\pi = 1.5$ MHz and $\zeta_2/2\pi = 3.5$ MHz. Other parameters are specified in the subsection \ref{xx}.}\label{ddd}
		\end{center}
	\end{figure}
	Figure~\ref{ddd} illustrates the group delay $\tau$ of the output probe field as a function of the normalized detuning $\delta/\omega_{d}$ for different interaction strengths. In Figure~\ref{ddd}(a), the photon-magnon coupling $\chi_1$ is varied while the remaining parameters are fixed at $\zeta_1/2\pi= \chi_2/2\pi=1.5$ MHz, $\zeta_2/2\pi=3.5$ MHz, and $\zeta_a/2\pi=2.5$ MHz. For $\chi_1/2\pi=0.5$ MHz, a negative group delay appears around $\delta \approx 0.9905\omega_d$, revealing the occurrence of a fast light regime. A positive group delay is also observed near $\delta \approx 0.9908\omega_d$, corresponding to a slow light regime in the same spectral region. As $\chi_1/2\pi$ is increased to $1.5$ MHz, both the positive and negative group-delay peaks become more pronounced, indicating that the stronger photon-magnon coupling enhances the dispersion responsible for the slow and fast light effects. A further increase to $\chi_1/2\pi=2.5$ MHz significantly enhances the slow light response around $\delta \approx 0.9915\omega_d$, where the group delay reaches a maximum value of about $\tau \approx 400~\mu\mathrm{s}$. In Figure~\ref{ddd}(b), we examine the influence of the photon-magnon coupling $\chi_2$ on the group delay $\tau$ of the output probe field, while the remaining parameters are kept fixed as indicated in the figure. For $\chi_2/2\pi=0.5$ MHz, two negative group-delay regions appear around $\delta \approx 0.769\omega_d$ and $\delta \approx 0.99081\omega_d$, corresponding to fast-light propagation of the output probe field. In addition, positive group delays are observed near $\delta \approx 0.99061\omega_d$, $\delta \approx 0.99105\omega_d$, and $\delta \approx 1.135\omega_d$, indicating slow-light propagation. As the coupling strength is increased to $\chi_2/2\pi=3$ MHz and then to $5$ MHz, the magnitudes of both the positive and negative group delays around $\delta \approx 0.99061\omega_d$, $\delta \approx 0.99081\omega_d$, and $\delta \approx 0.99105\omega_d$ gradually decrease. Meanwhile, the slow and fast light features located near $\delta \approx 1.135\omega_d$ and $\delta \approx 0.769\omega_d$, respectively, become progressively weaker and eventually approach zero group delay. In Figure~\ref{ddd}(c), we investigate the effect of the magnon-phonon coupling $\zeta_1$ on the group delay $\tau$ of the output probe field, while the remaining parameters are kept fixed. For $\zeta_1/2\pi=1$ MHz, both the positive and negative group-delay peaks are relatively small, indicating weak slow- and fast-light responses. As the coupling strength is increased to $\zeta_1/2\pi=1.5$ MHz, both group-delay peaks become more pronounced, indicating an enhancement of the slow- and fast-light responses. When $\zeta_1/2\pi$ is further increased to $2$ MHz, the slow-light effect becomes dominant. A pronounced positive group-delay peak appears around $\delta \approx 0.99\omega_d$, where the group delay reaches a maximum value of $\tau \approx 273.6~\mu\mathrm{s}$, while the negative group delay is considerably reduced. In Figure~\ref{ddd}(d), we examine the effect of the magnon-phonon coupling $\zeta_2$ on the group delay $\tau$ of the output probe field, while the remaining parameters are kept fixed. For $\zeta_2/2\pi=1$ MHz, positive group-delay peaks are observed at $\delta \approx 0.966\omega_d$, $\delta \approx 0.99255\omega_d$, $\delta \approx 0.99104\omega_d$, and $\delta \approx 1.033\omega_d$, whereas a negative group-delay peak appears at $\delta \approx 0.99079\omega_d$. As $\zeta_2$ is increased, the positive group-delay peaks gradually become smaller, while the negative group-delay peak becomes more pronounced. Finally, Figure~\ref{ddd}(e) illustrates the effect of the photon-phonon coupling strength $\zeta_a$ on the group delay $\tau$ of the output probe field, while all other parameters are kept fixed. For $\zeta_a/2\pi = 1$~MHz, a pronounced positive group-delay peak is observed around $\delta \approx 0.99\omega_d$, indicating a strong slow-light response. As $\zeta_a/2\pi$ is increased to $2$~MHz, this positive group-delay peak is significantly reduced. With a further increase to $\zeta_a/2\pi = 3$~MHz, the positive group delay decreases even further, while a negative group delay emerges, marking the transition to the fast light regime. The tunability of the group delay originates from the quantum interference between the different excitation pathways in the hybrid system, indicating that the propagation of the output probe field can be dynamically controlled by adjusting the photon-magnon, magnon-phonon, and photon-phonon coupling strengths.\\
	Figure~\ref{dddd} shows the group delay $\tau$ of the output probe field as a function of the normalized detuning for different values of the Kerr-induced frequency shift $\Delta K$. In the absence of the Kerr effect ($\Delta K=0$), the group delay remains positive over the considered detuning range, indicating a slow-light regime. As the Kerr-induced frequency shift is introduced, negative group-delay regions emerge, corresponding to the appearance of a fast-light regime. These results demonstrate that the Kerr-induced frequency shift provides an effective means of controlling the propagation characteristics of the output probe field by enabling a transition between the slow- and fast-light regimes.
	\begin{figure}
		\begin{center}
			\includegraphics[scale=0.6]{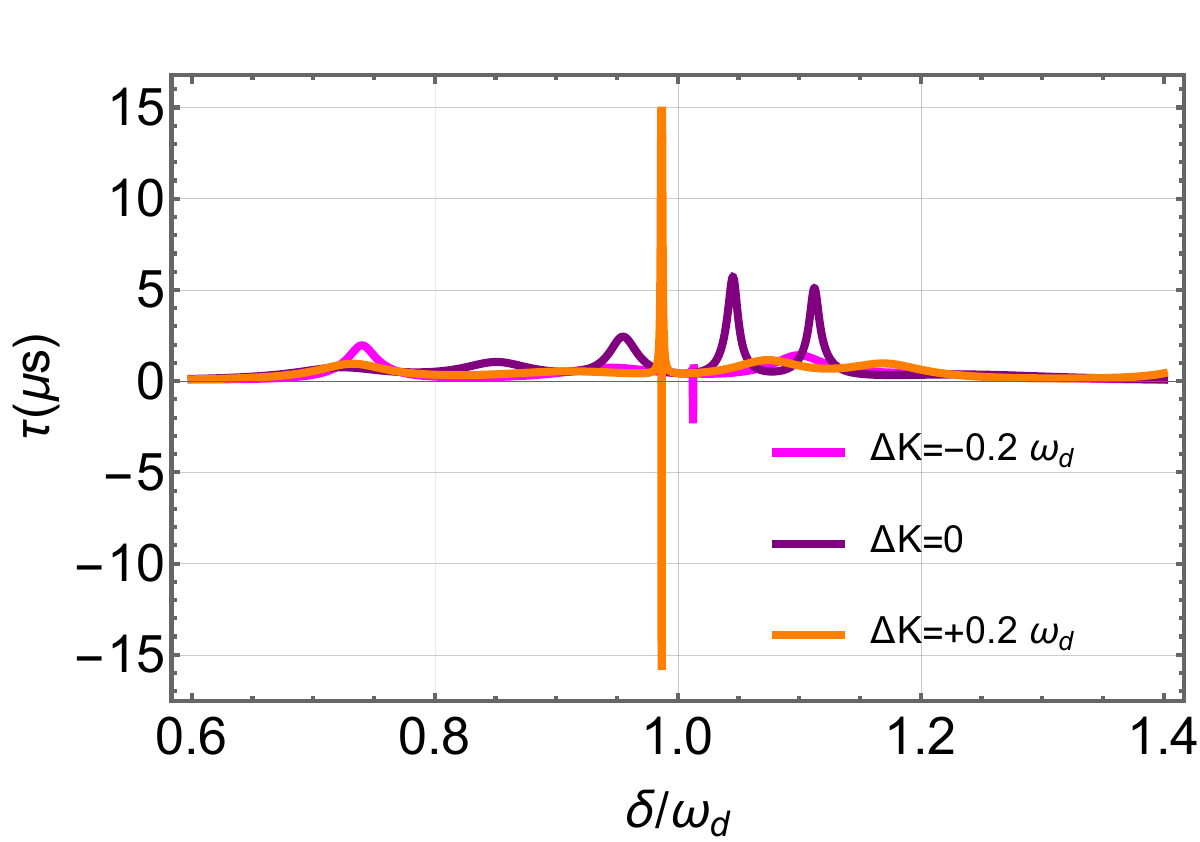}
			\caption{Group delay $\tau$ plotted versus the normalized detuning $\delta/\omega_d$ for different values of the Kerr effect $\Delta K$, while the other parameters are fixed as specified in subsection \ref{xx}.}\label{dddd}
		\end{center}
	\end{figure}
	\section{Nonreciprocity} \label{xxxx}
	In this section, we investigate the nonreciprocal behavior of the absorption spectrum and group delay induced by the Kerr effect, which plays an essential role in controlling directional signal propagation and facilitating the realization of integrated nonreciprocal photonic devices. Since the analytical expressions for the absorption spectrum $\text{Re}[\epsilon_{\mathrm{out}}]$ and the group delay $\tau$ have already been obtained in Eqs.~\eqref{E} and~\eqref{cv}, respectively, we directly study their dependence on the Kerr frequency shift $\Delta K$.
	\subsection{Nonreciprocal absorption}
	To characterize the Kerr-effect-induced nonreciprocal absorption, we define the contrast ratio $\epsilon_{NP}$ (with $0 \leq \epsilon_{NP} \leq 1$) as
	\begin{equation}
		\epsilon_{NP}=\frac{\left|\text{Re}[\epsilon_{\mathrm{out}}]\left(\Delta K<0\right)-\text{Re}[\epsilon_{\mathrm{out}}]\left(\Delta K>0\right)\right|}{\text{Re}[\epsilon_{\mathrm{out}}]\left(\Delta K<0\right)+\text{Re}[\epsilon_{\mathrm{out}}]\left(\Delta K>0\right)}.
	\end{equation}
	
	The condition $\epsilon_{NP}=0$ corresponds to reciprocal absorption, whereas $\epsilon_{NP}=1$ indicates perfect nonreciprocal absorption.
	\begin{figure} [h!] 
		\begin{center}
			\includegraphics[scale=0.44]{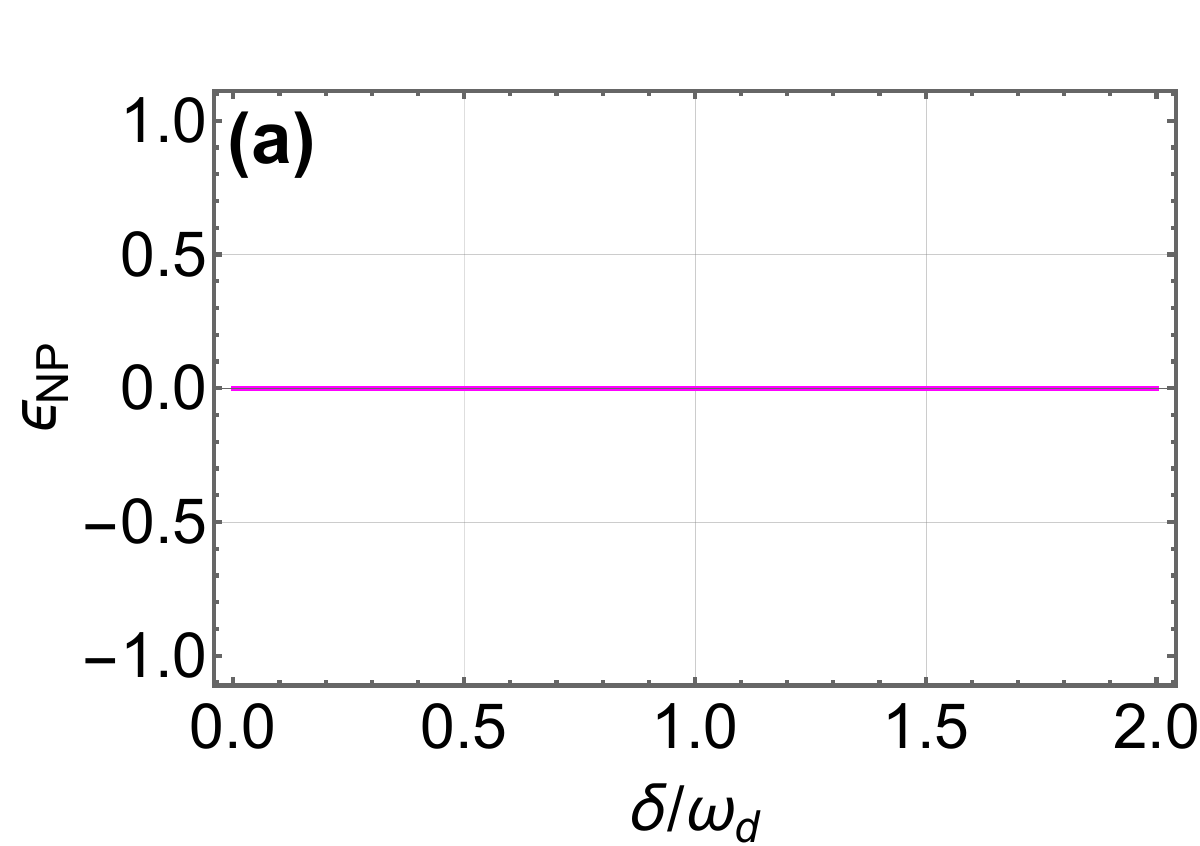}
			\includegraphics[scale=0.43]{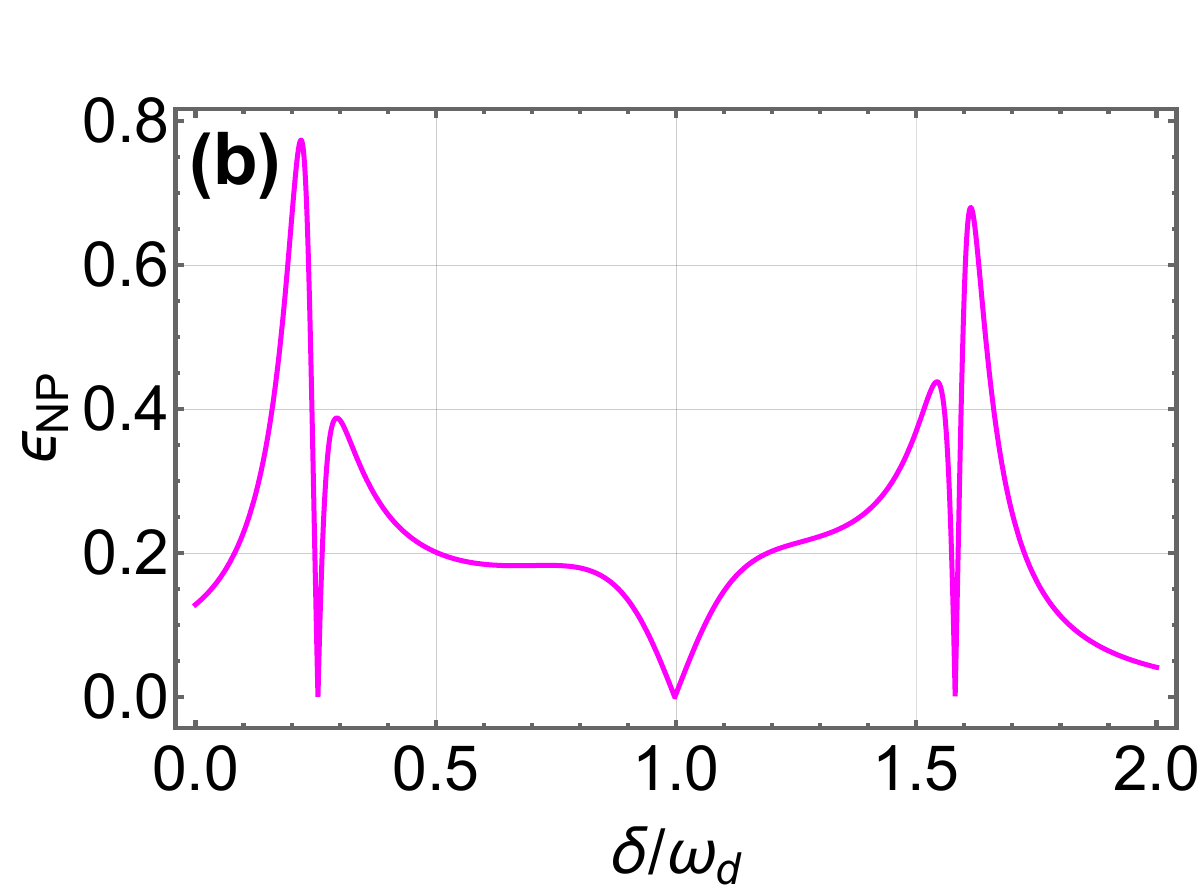}
			\includegraphics[scale=0.43]{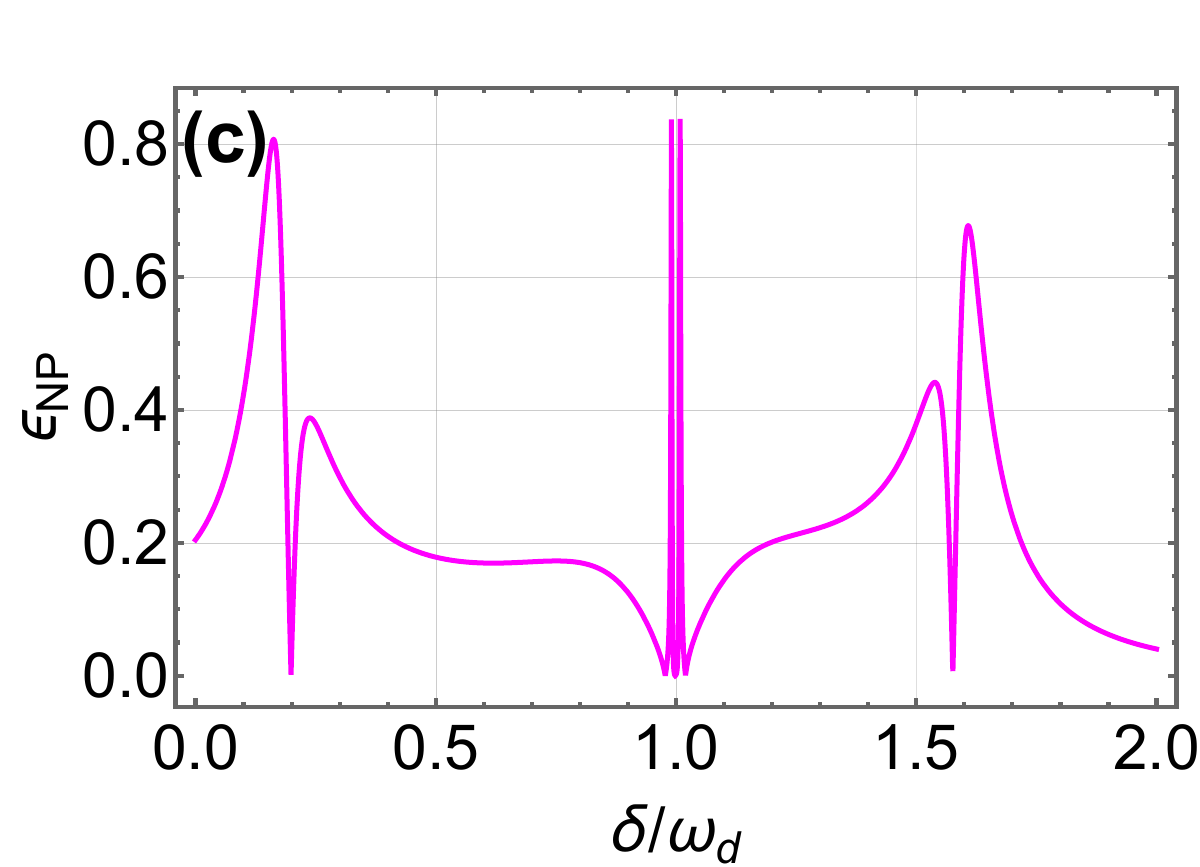}
			\includegraphics[scale=0.43]{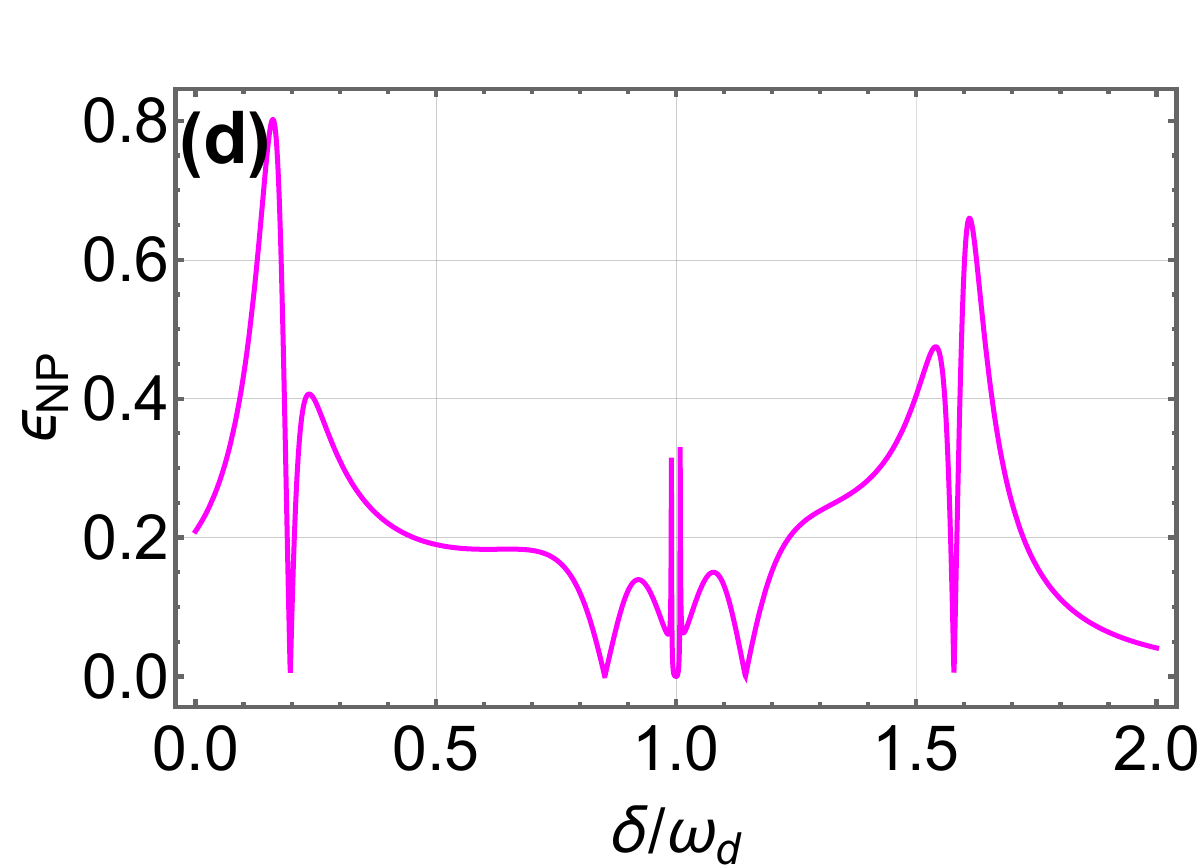}
			\includegraphics[scale=0.43]{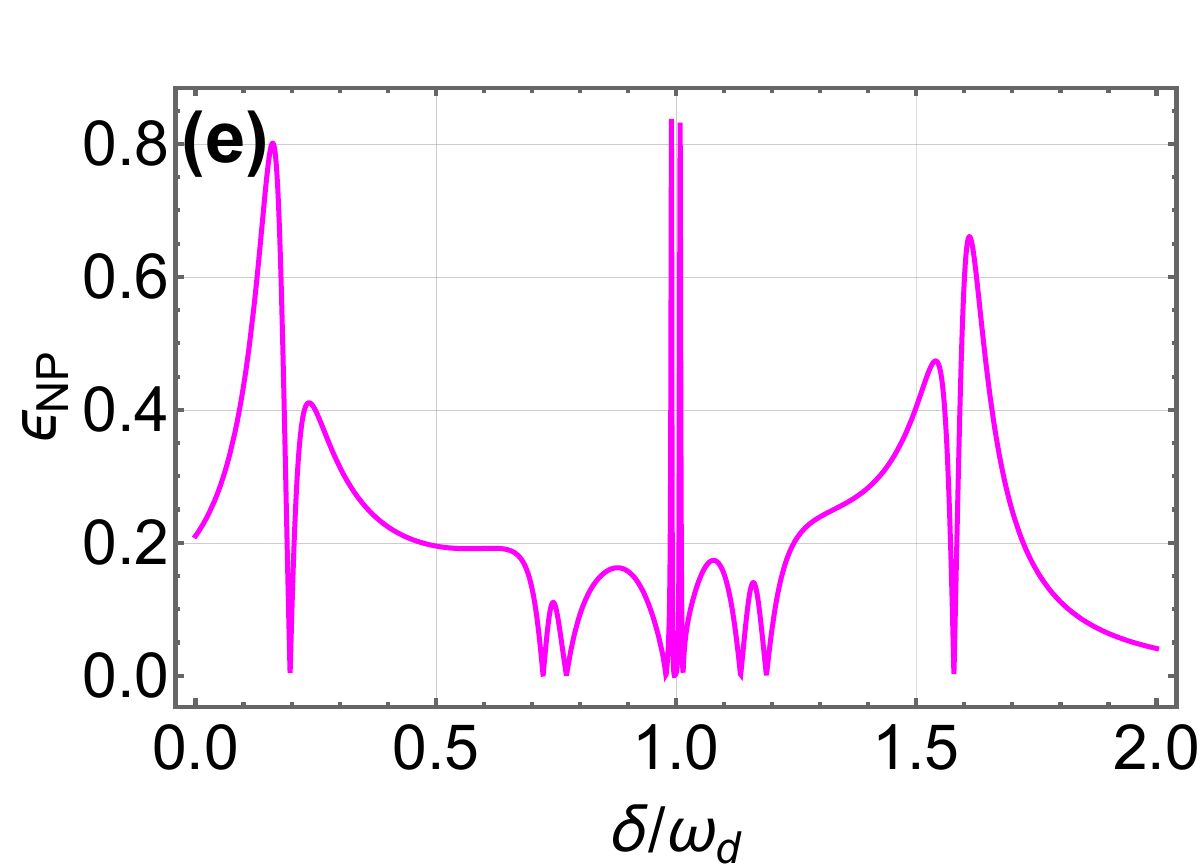}
			\includegraphics[scale=0.43]{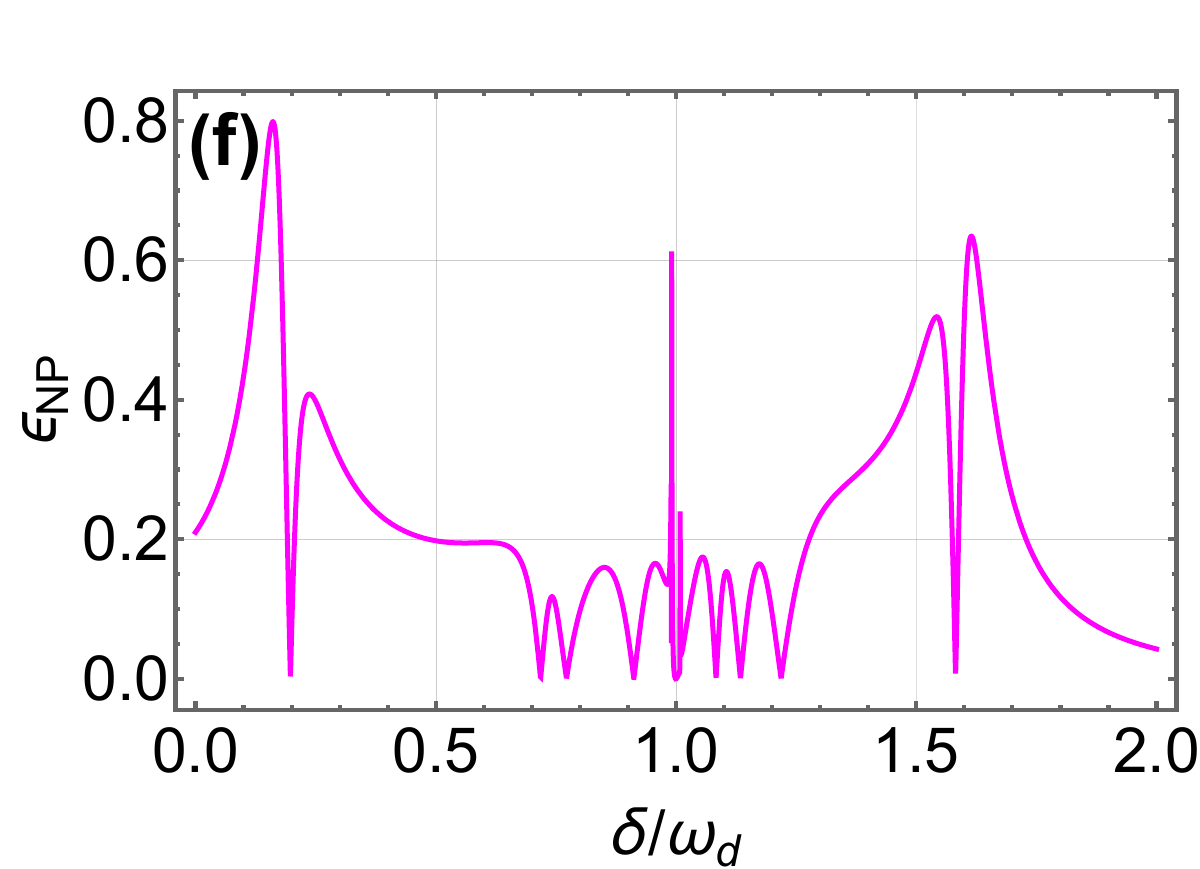}
			\caption{ Nonreciprocal absorption versus $\delta/\omega_d$, with $|\Delta K|=0.3\omega_d$. (a) $\chi_{1}=\chi_{2}=\zeta_{1}=\zeta_{2}=\zeta_a=0$; (b) $\chi_{1}/2\pi=1.5$ MHz, $\chi_{2}=\zeta_{1}=\zeta_{2}=\zeta_a=0$; (c) $\chi_{1}/2\pi=\zeta_{1}/2\pi=1.5$ MHz, $\chi_{2}=\zeta_{2}=\zeta_a=0$; (d) $\chi_{1}/2\pi=\chi_{2}/2\pi=\zeta_{1}/2\pi=1.5$ MHz, $\zeta_{2}=\zeta_a=0$; (e) $\chi_{1}/2\pi=\chi_{2}/2\pi=\zeta_{1}/2\pi=1.5$ MHz, $\zeta_{2}/2\pi=3.5$ MHz $\zeta_a=0$; and (f) $\chi_{1}/2\pi=\chi_{2}/2\pi=\zeta_{1}/2\pi=1.5$ MHz, $\zeta_{2}/2\pi=3.5$ MHz, $\zeta_a/2\pi=2.5$ MHz. Other parameters are as in Sec. \ref{xx}.}\label{s}
		\end{center}
	\end{figure} 
	Figure~\ref{s} demonstrates that the nonreciprocal absorption is fully controllable via the normalized detuning $\delta/\omega_d$, allowing reversible switching between reciprocal and nonreciprocal responses. The nonreciprocity contrast $\epsilon_{\mathrm{NP}}$ can be tuned continuously from 0 to 1 by adjusting $\delta/\omega_d$, providing efficient control of absorption asymmetry and enabling reconfigurable transparency windows. In Figure~\ref{s}(a), when all couplings are neglected, the nonreciprocity contrast remains zero, indicating the absence of nonreciprocal behavior. This is consistent with Figure~\ref{b}(a), where the curves for $\Delta K < 0$ and $\Delta K > 0$ overlap, confirming a fully reciprocal response. In Figure~\ref{s}(b), optimal nonreciprocity is observed for $0.078 < \delta/\omega_d < 0.24$ and $1.58 < \delta/\omega_d < 1.72$. In Figure~\ref{s}(c), maximum nonreciprocity occurs in narrow windows around resonance, specifically within the ranges $0.983 \le \delta/\omega_d \le 1.004$ and $1.005 \le \delta/\omega_d \le 1.015$, achieving $\epsilon_{\mathrm{NP}} = 0.99$. For Figures~\ref{s}(d) and~\ref{s}(f), the regions of maximum nonreciprocity are the same as those shown in Figure~\ref{s}(b), while those in Figure~\ref{s}(e) coincide with those of Figure~\ref{s}(c). By comparing Figures~\ref{s}(b)–\ref{s}(f) with Figures~\ref{b}(b)–\ref{b}(f), respectively, we observe that when the orange curve ($\Delta K > 0$) and the magenta curve ($\Delta K < 0$) overlap, the system exhibits no nonreciprocal response. In contrast, a clear separation between the two curves indicates strong nonreciprocal behavior. These results show that, by appropriately tuning the detuning $\delta$, the transparency windows and absorption spectra associated with opposite Kerr configurations become increasingly asymmetric, demonstrating that the Kerr-induced nonreciprocal response can be efficiently controlled and tuned.\\
	\subsection{Nonreciprocal group delay}
	
\begin{figure} [h!] 
		\begin{center}
			\includegraphics[scale=0.5]{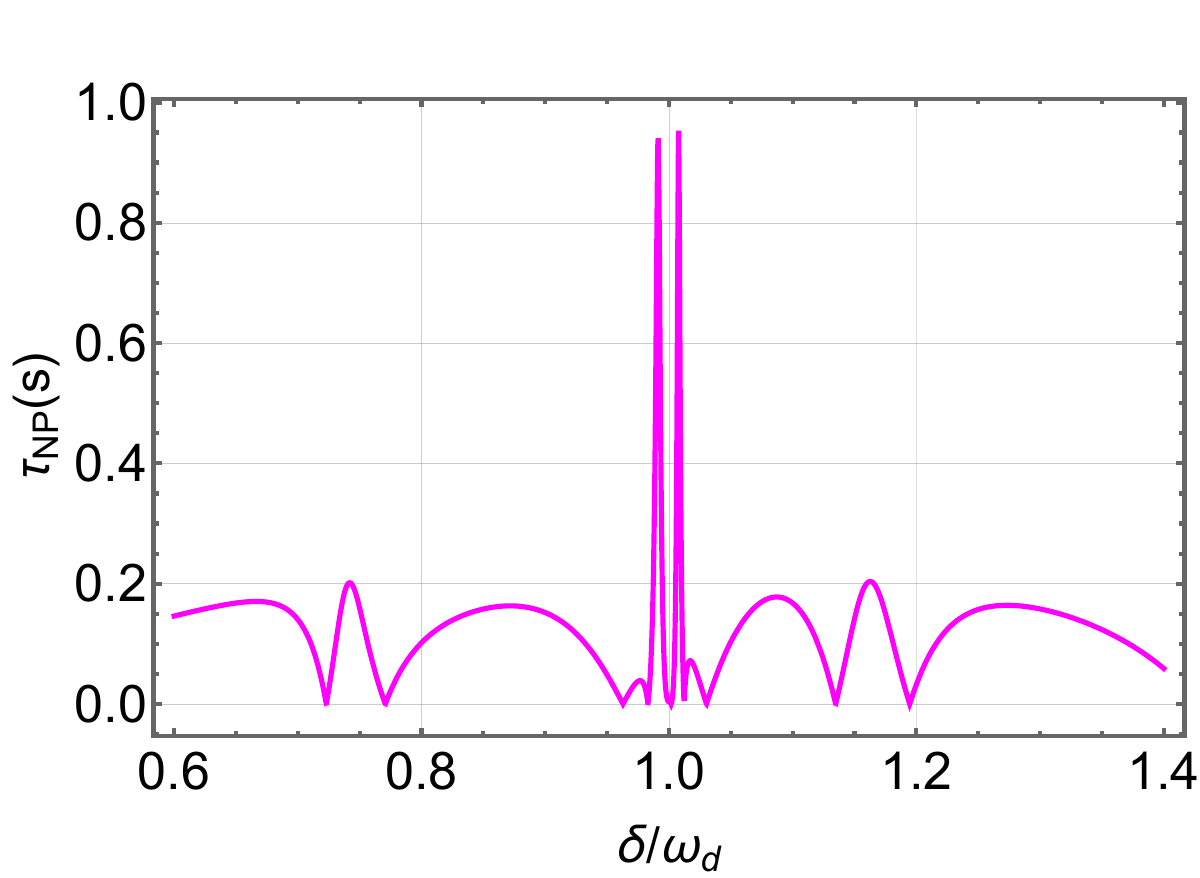}
			\caption{ Nonreciprocal group delay as a function of $\delta/\omega_{d}$, with $ \zeta_{a}/2\pi = 1$ MHz and $|\Delta K|=0.3\omega_{d}$. All remaining parameters are given in subsection \ref{xx}.}\label{s8}
		\end{center}
	\end{figure}

	We now investigate the directional behavior of the group delay, whose analytical form is given in Eq.~\eqref{cv}. Since the phase dispersion of the transmitted field is strongly influenced by the hybrid photon–magnon–phonon interactions, reversing the sign of $\Delta K$ modifies the phase slope and consequently the propagation time. To quantify this directional asymmetry, we introduce the group-delay nonreciprocity factor as
	
	\begin{equation}
		\tau_{NP}=\frac{\left|\tau\left(\Delta K<0\right)-\tau\left(\Delta K>0\right)\right|}{\tau\left(\Delta K<0\right)+\tau\left(\Delta K>0\right)},
	\end{equation}
	
	where $\tau_{NP}=0$ corresponds to fully reciprocal propagation, while $\tau_{NP}=1$ represents ideal nonreciprocal group delay. As shown in Figure~\ref{s8}, the nonreciprocal response of the group delay can be effectively controlled by tuning the normalized detuning $\delta/\omega_d$. In particular, the nonreciprocity contrast can be continuously adjusted from 0 to 1 by varying $\delta/\omega_d$. Strong nonreciprocal behavior is observed within the narrow ranges $0.97 \le \delta/\omega_d \le 0.99$ and $1 \le \delta/\omega_d \le 1.01$, where the group delay exhibits a pronounced asymmetry for opposite Kerr-induced frequency shifts. Physically, this means that the transmitted probe field experiences different propagation delays depending on the sign of the Kerr-induced frequency shift, enabling directional control of the slow- and fast-light responses.
	\section{Feasibility}\label{Fea}
	The experimental feasibility of the proposed model is supported by significant advances in cavity magnomechanical and optomechanical systems \cite{0,24,54,55}. Experimental studies have demonstrated that magnomechanically induced transparency originates from coherent interference arising from the strong coupling between magnon and phonon modes \cite{0}. In particular, collective magnon excitations in yttrium iron garnet (YIG) spheres embedded in microwave cavities exhibit strong coupling to microwave photons via magnetic dipole interaction \cite{25}, while magnetostrictive effects provide an efficient coupling mechanism between magnons and phonons \cite{24}. In addition, optomechanical interactions can be realized using radiation-pressure coupling, for example by integrating a high-reflectivity mirror with a micro-mechanical resonator \cite{54}. Such systems have already demonstrated optomechanically induced transparency under realistic experimental conditions \cite{55}. Moreover, the Kerr nonlinearity in YIG, originating from magnetocrystalline anisotropy, has been experimentally observed under strong microwave driving and has been widely exploited to investigate nonlinear magnon dynamics, bistability, and frequency tuning \cite{45,56}. Based on these established techniques, coherent interactions among magnons, phonons, and photons, together with Kerr-induced nonlinear effects, can be effectively implemented \cite{45,56}. Importantly, in realistic experimental conditions, finite temperature leads to nonzero thermal populations of magnons and phonons, which introduce additional fluctuations into the system \cite{0}. These thermal effects can reduce the visibility and contrast of the transparency windows, as well as the magnitude of the group delay and the nonreciprocal response. However, current experiments routinely operate at cryogenic temperatures, where thermal excitations are significantly suppressed, allowing the coherent features predicted in this work to remain observable \cite{54,Genes2008}. Furthermore, typical experimental parameters, such as strong magnon-photon coupling rates, low mechanical damping, operation in the resolved-sideband regime, and experimentally accessible Kerr nonlinearities, are all within reach of current state-of-the-art platforms \cite{0,18,24,56}. Therefore, since all essential ingredients of the proposed hybrid system have already been demonstrated individually, the present scheme can be realistically implemented with existing experimental technology.
	\section{Conclusion}\label{xxxxx}
	
	In this work, we have theoretically investigated the optical response of a hybrid cavity magnomechanical system composed of two YIG spheres and a mechanical membrane coupled to a microwave cavity in the presence of magnon Kerr nonlinearity. Using the linearized quantum Langevin approach together with the input-output formalism, we analyzed the transmission properties of a weak probe field and explored the interplay among cavity photons, magnons, and mechanical vibrations. Our results reveal the emergence of multiple transparency windows arising from different coherent interference mechanisms within the hybrid platform. In particular, the photon-magnon interaction generates magnon-induced transparency, while the magnomechanical interactions give rise to multiple magnomechanically induced transparency windows. The additional photon-phonon coupling further enriches the spectral response and produces optomechanically induced transparency. The coexistence of these interaction channels leads to strong multimode interference and a highly tunable transmission spectrum. We have shown that the magnon Kerr nonlinearity plays a central role in controlling the spectral properties of the system. The Kerr-induced nonlinear frequency shift modifies the effective magnon resonance and consequently reshapes the interference pathways among the hybrid modes. As a result, the transparency windows can be shifted and distorted, leading to the appearance of asymmetric Fano-like resonances. Furthermore, the cavity and magnon dissipation rates strongly influence the visibility and linewidth of the transparency features, providing additional control over the system response. We have also investigated the dispersive properties of the transmitted probe field through the associated group delay. Depending on the hybrid coupling strengths and Kerr-induced frequency shift, the system exhibits both slow- and fast-light propagation regimes. The group delay can be significantly enhanced or suppressed by tuning the photon-magnon and magnon-phonon interactions, demonstrating a flexible mechanism for controlling microwave signal propagation and storage. In addition, we demonstrated that the Kerr nonlinearity enables highly tunable nonreciprocal absorption and group delay, allowing reversible switching between reciprocal and nonreciprocal propagation by controlling the Kerr-induced frequency shift. Overall, our study demonstrates that Kerr-engineered cavity magnomechanical systems constitute a versatile platform for manipulating transparency, interference phenomena, and light propagation in the microwave domain. The ability to control multiple transparency windows, generate tunable Fano resonances, and switch between slow- and fast-light regimes may find promising applications in microwave photonics, coherent signal processing, quantum communication, and hybrid quantum information technologies.

	\appendix
	\renewcommand{\thesection}{ \Alph{section}}
	\section{Derivation of $c_-$} \label{AP}
	\begin{align}
		\alpha_1 &= \kappa_c + i(\bar{\Delta}_c - \delta), &
		\alpha_2 &= \kappa_{n_1} + i(\bar{\Delta}_{n_1} + 2\Delta K - \delta), &
		\alpha_3 &= \omega_{d_1} - \frac{\delta}{\omega_{d_1}}(\delta + i \gamma_{d_1}) \\
		\alpha_4 &= \kappa_{n_1} - i(\bar{\Delta}_{n_1} + 2\Delta K + \delta), &
		\alpha_5 &= \kappa_c - i(\bar{\Delta}_c + \delta), &
		\alpha_6 &= \omega_{d_3} - \frac{\delta}{\omega_{d_3}}(\delta + i \gamma_{d_3}) \\
		\alpha_7 &= \kappa_{n_2} - i(\bar{\Delta}_{n_2} + \delta), &
		\alpha_8 &= \omega_{d_2} - \frac{\delta}{\omega_{d_2}}(\delta + i \gamma_{d_2}), &
		\alpha_9 &= \kappa_{n_2} + i(\bar{\Delta}_{n_2} - \delta)\\
		\mathcal{A} &= 1 + \frac{\mathcal{G}_2^2}{i \alpha_8 \alpha_9}, &
		\mathcal{B} &= 1 - \frac{\mathcal{G}_2^2}{i \alpha_7 \alpha_8 \mathcal{A}}, &
		\mathcal{C} &= 1 + \frac{\chi_2^2}{\alpha_5 \alpha_7 \mathcal{B}} + \frac{\mathcal{G}_c^2}{i \alpha_5 \alpha_6} \\
		\mathcal{D} &= \frac{i \chi_2^2 \mathcal{G}_2^2}{\alpha_5 \alpha_7 \alpha_8 \alpha_9 \mathcal{A} \mathcal{B}} + \frac{\mathcal{G}_c^2}{i \alpha_5 \alpha_6}, &
		\mathcal{E} &= 1 + \frac{\chi_1^2}{\alpha_4 \alpha_5 \mathcal{C}} - \frac{\mathcal{G}_1^2}{i \alpha_3 \alpha_4}, &
		\mathcal{F} &= \frac{\mathcal{G}_1^2}{i \alpha_3 \alpha_4} + \frac{i \Delta K}{\alpha_4} \\
		\mathcal{G} &= 1 + \frac{\mathcal{G}_1^2}{i \alpha_2 \alpha_3}, &
		\mathcal{H} &= \frac{\mathcal{G}_1^2}{i \alpha_2 \alpha_3} + \frac{i \Delta K}{\alpha_2}, &
		\mathcal{I} &= 1 + \frac{\mathcal{H} \mathcal{F}}{\mathcal{G} \mathcal{E}} \\
		\mathcal{J} &= \frac{-i \chi_1}{\alpha_2 \mathcal{G}} + \frac{i \chi_1 \mathcal{H} \mathcal{D}}{\alpha_4 \mathcal{C} \mathcal{E} \mathcal{G}}&
		\mathcal{K} &= 1 - \frac{\mathcal{G}_1^2}{i \alpha_3 \alpha_4}, &
		\mathcal{L} &= \frac{\mathcal{G}_1^2}{i \alpha_3 \alpha_4} + \frac{i \Delta K}{\alpha_4},\\
		\mathcal{M} &= 1 + \frac{\mathcal{H} \mathcal{L}}{\mathcal{G} \mathcal{K}}  &
		\mathcal{N} &= 1 + \frac{\chi_1^2}{\alpha_4 \alpha_5 \mathcal{K} \mathcal{M}} + \frac{\mathcal{G}_c^2}{i \alpha_5 \alpha_6}, &
		\mathcal{O} &= \frac{-\chi_1^2 \mathcal{L}}{\alpha_2 \alpha_5 \mathcal{K} \mathcal{M} \mathcal{G}} + \frac{\mathcal{G}_c^2}{i \alpha_5 \alpha_6}, \\
		\mathcal{P} &= 1 + \frac{\chi_2^2}{\alpha_7 \alpha_5 \mathcal{N}} &
		\mathcal{Q} &= 1 - \frac{\mathcal{G}_2^2}{i \alpha_7 \alpha_8 \mathcal{P}}, &
		\mathcal{R} &= 1 + \frac{\mathcal{G}_2^2}{i \alpha_8 \alpha_9 \mathcal{Q}}, \\
		\mathcal{S} &= \frac{-i \chi_2}{\alpha_9} + \frac{\chi_2 \mathcal{G}_2^2 \mathcal{O}}{\alpha_7 \alpha_8 \alpha_9 \mathcal{N} \mathcal{P} \mathcal{Q}} &
		\mathcal{T} &= 1 + \frac{\chi_1^2}{\alpha_4 \alpha_5 \mathcal{K} \mathcal{M}} + \frac{\chi_2^2}{\alpha_5 \alpha_7 \mathcal{B}}, &
		\mathcal{U} &= \frac{-\chi_1^2 \mathcal{L}}{\alpha_2 \alpha_5 \mathcal{K} \mathcal{M} \mathcal{G}} + \frac{i \chi_2^2 \mathcal{G}_2^2}{\alpha_5 \alpha_7 \alpha_8 \alpha_9 \mathcal{A} \mathcal{B}} \\
		\mathcal{V} &= 1 + \frac{\mathcal{G}_c^2}{i \alpha_5 \alpha_6 \mathcal{T}}, &
		\mathcal{W} &= \frac{-\mathcal{G}_c}{i \alpha_6} + \frac{\mathcal{G}_c \mathcal{U}}{i \alpha_6 \mathcal{T}},
	\end{align}
	where, $\mathcal{G}_{1}=\zeta_1/\sqrt{2}$ $\left(\mathcal{G}_{2}=\zeta_2/\sqrt{2}\right)$ and $\mathcal{G}_{c}=\zeta/\sqrt{2}$, with ${\zeta}_1=i\sqrt{2}\zeta_{01}n_{1s}$ $\left({\zeta}_2=i\sqrt{2}\zeta_{02}n_{2s}\right)$ and $\zeta=i\sqrt{2}\chi_cc_{s}$, representing the effective opto-magnomechanical coupling rate, where $|\bar{\Delta}_{n_1}|,|\bar{\Delta}_{n_2}|,\left|\bar{\Delta}_{c} \right| \gg \kappa_{n_1}, \kappa_{n_2},\kappa_{c}$.\\

	\section*{Acknowledgments}
	
	M. Amghar expresses gratitude for the financial support he receives from the National Center for Scientific and Technical Research (CNRST) under the “PhD-Associate Scholarship-PASS” program. This work was supported by Princess Nourah bint Abdulrahman University Researchers Supporting Project number (PNURSP2026R399), Princess Nourah bint Abdulrahman University, Riyadh, Saudi Arabia. The authors are thankful to the Deanship of Graduate Studies and Scientific Research at University of Bisha for supporting this work through the Fast-Track Research Support Program.

\end{document}